\documentclass[aps,pra,superscriptaddress,twocolumn]{revtex4-2}
\usepackage{graphicx,amsmath,amssymb,amsfonts,latexsym,color,dcolumn,bm}

\newcommand{\beq}{\begin{equation}}
\newcommand{\eeq}{\end{equation}}
\newcommand{\bea}{\begin{eqnarray}}
\newcommand{\eea}{\end{eqnarray}}

\providecommand{\bra}[1]{\langle #1 \rvert}
\providecommand{\ket}[1]{\lvert #1 \rangle}

\usepackage{amsfonts,amssymb}
\usepackage{dsfont}
\usepackage{physics}
\usepackage{tocvsec2}

\usepackage{appendix}

\usepackage{tikz}

\usepackage{soul} 

\usepackage{amsmath}
\usepackage{bbold}
\usepackage{hyperref}
\hypersetup{
     colorlinks=true,      
    linkcolor=blue,        
    citecolor=blue,        
    filecolor=magenta, 
    urlcolor=black          
}

\usepackage{stmaryrd} 

\usepackage{comment}

\definecolor{2ca25f}{HTML}{2ca25f}
\definecolor{99d8c9}{HTML}{99d8c9}
\definecolor{e5f5f9}{HTML}{E5F5F9}

\definecolor{9ebcda}{HTML}{9EBCDA}
\definecolor{8856a7}{HTML}{8856a7}

\definecolor{fee8c8}{HTML}{fee8c8}
\definecolor{e34a33}{HTML}{e34a33}

\begin{document}
\author{Mohamed Meguebel}
\email{mohamed.meguebel@telecom-paris.fr}
\affiliation{Telecom Paris, Institut Polytechnique de Paris, 19 Place Marguerite Perey, 91120 Palaiseau, France}
\author{Nadia Belabas}
\affiliation{Centre for Nanosciences and Nanotechnology, CNRS, Université Paris-Saclay, 10 Bd Thomas Gobert, 91120 Palaiseau, France}
\author{Nicolas Fabre}
\email{nicolas.fabre@telecom-paris.fr}
\affiliation{Telecom Paris, Institut Polytechnique de Paris, 19 Place Marguerite Perey, 91120 Palaiseau, France}

\date{\today}
\begin{abstract}
Quantum teleportation, a key building block for quantum technologies, enables distant parties, Alice and Bob, to transfer an unknown quantum state using entanglement, local operations, and classical communication. In this work, we propose a quantum teleportation protocol for single-photon time-frequency continuous variables, where time is the time of arrival at a detector. We detail the operational requirements of the protocol, notably the nonlinear interactions needed for the photon-photon time-frequency entangling gate. By explicitly treating the detector position as the reference frame for time of arrival, we show that, even with ideal classical and quantum channels, the input-output fidelity is not unity unless the receiving party can correct the detector-position mismatch. This results in a trade-off between teleportation fidelity and the temporal bandwidth of the input photon wavepacket. The protocol itself can be used for remote detector synchronization. In particular, we show that time-frequency cat and Gottesman-Kitaev-Preskill states provide an internal temporal ruler through fidelity revivals. We further derive the classical and best-copy fidelity thresholds for single-photon Gaussian wavepackets, finding that the relevant vacuum-equivalent shot noise is input-dependent, in contrast with quadrature-based continuous-variable quantum teleportation. Our proposal derives from a quantum theory of space-time events, whereafter we establish space as a reference for the time degree of freedom, thereby grounding our work in ongoing research on reference frames and superselection rules in quantum optics and quantum information. Beyond protocols harnessing time-frequency continuous variables of single photons, this work points toward remote synchronization in quantum communication architectures.
\end{abstract}
\pacs{}
\vskip2pc

\title{Time-Frequency Continuous-Variable Quantum Teleportation with Single Photons}
\maketitle
\section{Introduction}
Quantum information is the bedrock of second-revolution quantum technologies~\cite{ContextQtechEuropeanacin2018quantum}. In particular, photonic quantum technologies~\cite{ContextphotonicobrienFurusawa2009photonic,Contextslussarenko2019photonic} employ photon-based quantum information encodings which are often broken down into two categories: discrete variables (DVs)~\cite{DVQIjaeger2007quantum} and continuous variables (CVs)~\cite{ContextCVserafini2023quantum}. The quantized electromagnetic quadratures~\cite{CVfieldquadraturebraunstein2005quantum} with their associated canonical commutation relation (CCR) $[\hat x, \hat p] = i$ arguably constitute the most prevalent CV encoding in quantum optics, exploiting the photon-number statistical distribution of a given mode as the carrier of quantum information. In recent years, however, other CV encodings with their corresponding CCRs, such as single-photon transverse position-momentum~\cite{tasca_continuous_2011} and time-frequency (TF) CVs~\cite{TFCVfabre2022time}, have been gaining traction. In particular, TF CVs and their discrete-mode analogues~\cite{lukens2016frequency,brecht2015photon} exploit single-photon frequency and time-of-arrival degrees of freedom, benefiting from their resilience against linear fiber impairments, such as polarization mode dispersion, and from the versatility of programmable filters~\cite{henry2024parallelization} to realize single-photon gates. Single-photon TF CVs have been investigated in a handful of quantum information schemes, such as quantum metrology~\cite{fabre2021parameter,descamps2023quantum,descamps_time-frequency_2023} and quantum error correction~\cite{fabre2020generation,yamazaki_linear_2023,fabre_teleportation-based_2023,descamps_gottesman-kitaev-preskill_2024}. \\
Quantum teleportation~\cite{Generalpirandola2015advances,Generalhu2023progress} -- consisting of a party, Alice, wishing to teleport an unknown state to another remote party, Bob, harnessing a shared entangled state -- is a prominent example of how quantum correlations can be employed in quantum technologies, underpinning for example quantum repeaters~\cite{briegel1998quantum} or measurement-based quantum computing (MBQC)~\cite{raussendorf2001one}. However, while it has been extensively investigated in more conventional photonic encodings such as quadrature CVs or polarization DVs, quantum teleportation has yet to be scrutinized for single-photon TF CVs.  While TF quantum teleportation was considered in~\cite{fabre_teleportation-based_2023} for quantum error correction, that work considered logical encodings obtained by discretizing CVs and relied on linear-optics Bell measurements. Humble proposed instead a single-photon spectral teleportation protocol~\cite{humble_spectral_2010}, employing nonlinear optics at the measurement stage, and thus manipulated frequency as a genuinely continuous degree of freedom. Nonetheless, the protocol in~\cite{humble_spectral_2010} is restricted to frequency, without considering the time of arrival or formulating time and frequency as CVs obeying the associated CCR $[\hat t,\hat \omega] = i$~\footnote{The authors in~\cite{TFCVfabre2022time} use the opposite sign convention $[\hat \omega,\hat t] = i$, consistent with their definition of the Fourier transform connecting the bosonic operators $\hat a(t)$ and $\hat a(\omega)$. These conventions are physically equivalent.} in the single-photon subspace~\cite{TFCVfabre2022time}. Such a TF CV description offers several advantages. It facilitates the derivation of general quantitative metrics such as input-output fidelities using phase-space methods~\cite{ContextCVserafini2023quantum}, particularly for Gaussian states~\cite{weedbrook2012gaussian,brask2021gaussian}. Moreover, it provides a practical framework for deriving quantum-teleportation benchmarks~\cite{braunstein2000criteria,grosshans2001quantum} and for establishing connections with other CV encodings, such as those based on electromagnetic-field quadratures. \\
Examining the quantum teleportation of a single-photon TF spectrum naturally prompts the question "What information is to be teleported?". More specifically, the time-of-arrival degree of freedom is intrinsically linked to the time operator, whose very definition has been debated since at least Pauli's time~\cite{pauliObjection2012general,Quantumclockleon2017pauli,Quantumclockreviewaltaie2022time}. Different proposals have been investigated, considering for example massive systems~\cite{zych2011quantum,roncallo2023does} or the use of a quantum clock~\cite{Quantumclockgiovannetti2015quantum,Quantumclockambiguimarletto2017evolution,Quantumclockmoreva2017quantum,Quantumclocksmith2019quantizing,Quantumclockmaccone2020quantum,Quantumclockfavalli2020time,Quantumcloclfoti2021time,Quantumclockreviewaltaie2022time,Quantumclockkuypers2025measuring}. The quantum clock description has attracted significant attention and can be summarized by the following sentence: \textit{time in a timeless universe}. This description is founded upon the Page-Wootters mechanism~\cite{pagewootters1983evolution} which, instead of treating time as an external parameter, regards it as an internal parameter of an enlarged system representing the global universe. Nonetheless, within the quantum clock proposal, time remains a conditioning variable rather than the quantum degree of freedom of primary interest. Furthermore, the framework requires the introduction of a quantum clock. Such an auxiliary system is not trivially instantiated in a concrete physical system and is not particularly appropriate for single-photon TF CV quantum teleportation, where the photon itself constitutes the quantum system carrying the information to be teleported and manipulated. Recently, Giovannetti \textit{et al.}~\cite{Eventbasedgiovannetti2023geometric} constructed a geometric event-based (GEB) quantum theory – subsuming the quantum clock description of time – in which space-time events are the primitive object. \\ \\
In this work, we propose a quantum teleportation protocol for single-photon TF CVs in which, building on the GEB description, the time degree of freedom is introduced through a detector-conditioned Page-Wootters mechanism. Our main findings are the following. (i) We show that the protocol teleports not only the photon's wavefunction but also the detector reference with respect to which it is defined. Consequently,  the resulting input-output state fidelity is not unity whenever there is a detector-position mismatch, even for ideal classical and quantum channels. In the present work, detector-position mismatch refers to any residual position-reference mismatch that cannot be corrected at the receiving end due to the absence of a common position reference between Alice and Bob. (ii) Accordingly, the teleportation fidelity becomes an experimental witness of both the classical and quantum channels' quality \textit{and} the synchronization of the detector references. This manifests as a fidelity/temporal-bandwidth trade-off: the narrower the photon in time, the more sensitive the fidelity is to detector-position mismatch. (iii) The detector-position mismatch can be determined and corrected within the teleportation protocol itself. By teleporting a known input state and optimizing the correction applied to the output, Bob can infer the detector-position mismatch and calibrate his detector reference relative to Alice's prior to teleporting the unknown target state. (iv) Finally, we derive two quantumness benchmarks for single-photon Gaussian wavepacket inputs: a classical-limit fidelity threshold $\max \mathcal F_{\text{class}}=1/2$ for measure-and-prepare strategies, and a best-copy fidelity threshold $\max \mathcal F_{1\rightarrow2}=2/3$, both identical to those computed for coherent state inputs in quadrature-based CV quantum teleportation~\cite{braunstein2000criteria,grosshans2001quantum}. Nevertheless, for both fidelity thresholds, we show that the input-independent vacuum shot noise from quadrature CVs is replaced by an input-dependent vacuum-equivalent shot noise for single-photon TF CVs.
\\ \\
The manuscript is organized as follows. In Sec.~\ref{Section Geometric event-based quantum mechanics}, we briefly introduce the GEB description of quantum mechanics, which we then apply to the light-cone constraint for photons in Sec.~\ref{Section Light-cone constraint}. In Sec.~\ref{Space-time foliation: detector as a reference}, we present our detector-conditioned Page-Wootters construction that sustains the rest of the work, where time is defined \textit{conditionally} on a detector reading. In Secs.~\ref{Section protocol Single- and two-photon states} and~\ref{Section protocol protocol steps}, we introduce the single-photon TF CV teleportation protocol and discuss its operational implementation, with an emphasis on the use of nonlinearities for the Bell measurement stage. In Sec.~\ref{Section Input-output state fidelity}, we derive the input-output fidelity for an ideal two-photon quantum channel and show how detector-position mismatch leads to a fidelity-bandwidth trade-off. We further demonstrate that TF cat and Gottesman-Kitaev-Preskill (GKP) states encode internal temporal references that give rise to fidelity revivals and can be exploited for remote detector-position synchronization. In Sec.~\ref{Section Fidelity threshold benchmarks}, we establish fidelity benchmarks for single-photon Gaussian wavepackets by considering, respectively, measure-and-prepare strategies and $1\rightarrow2$ Gaussian cloning, thereby identifying the conditions under which the teleportation protocol demonstrates a genuine quantum benefit. Finally, in Sec.~\ref{Section Discussion}, we discuss what sets this protocol apart from quadrature-based CV teleportation: (1) the physical unit underpinning the single-photon CV encoding, (2) the explicit treatment of reference frames that follows from defining time as the time of arrival at a detector, (3) the absence of an intrinsic vacuum noise. Sec.~\ref{Section Conclusion} concludes and gives perspectives, in particular for remote synchronization in quantum communication architectures. \\ 
Appendix~\ref{Appendix Protocol calculations in chronocyclic phase space} provides a general expression of the teleported output state by means of a derivation in chronocyclic phase space~\cite{BrechtPhysRevA.87.053810,fabre2020generation,TFCVfabre2022time}. A formula for the average input-output fidelity for a pure Gaussian input state and a general Gaussian two-photon quantum channel is computed. The TF correlations of the two-photon quantum channel required for surpassing the classical fidelity threshold $\max \mathcal F_\text{class} = 1/2$ are discussed. In Appendix~\ref{Appendix Section Superselection rules and reference externalization} we connect our detector-conditioned Page-Wootters construction to superselection rules (SSR) in quantum optics~\cite{Referenceframebartlett2007reference,Superselectionrulesdescamps2024superselection} and quantum reference frames.  \\ 
A supplementary material (Sup. Mat.) gives additional calculation details.

\section{Space-time event quantum optics}
\label{Space-time event quantum optics}
\subsection{Geometric Event-Based Quantum Mechanics}
\label{Section Geometric event-based quantum mechanics}
The GEB description of quantum mechanics introduced in~\cite{Eventbasedgiovannetti2023geometric} is underpinned by a kinematic Hilbert space $\mathcal H_{\text{GEB}}$ equipped with a basis $\{\ket{\overline x}\}$, where $\overline x$ is a four-dimensional space-time event $\overline x \equiv (x^0 =t, \vec x)$ and with $\langle \overline x | \overline x' \rangle = \delta^{(4)}(\overline x-\overline x')$. Considering a single spatial dimension, a physical state $\ket{\psi}_{\text{GEB}}$ of $\mathcal H_{\text{GEB}}$ can then be decomposed as
\begin{equation}
\label{Equation physical state in the GEB description}
\begin{split}
    \ket{\psi}_{\text{GEB}} = \int dtdx\; \psi_{\text{GEB}}(t,x)\ket{t,x}_{\text{GEB}}.
\end{split}
\end{equation}
This 1D reduction is motivated, for instance, when considering waveguide systems~\cite{roy2017colloquium} where the electromagnetic transverse modes are spatially confined. In GEB quantum mechanics, $\psi_{\text{GEB}}(t,x)$ is interpreted as the probability amplitude
that the event happens at space-time position $(t,x)$, that is at a given position $x$ \textit{and} time $t$. This signals that the GEB framework allows for a description where position and time are treated on an equal footing.

\subsection{Light-Cone Constraint}
\label{Section Light-cone constraint}
For photons propagating in vacuum, the light-cone constraint from special relativity~\cite{Specialrelativitysusskind2017special} imposes that the space-time interval $\Delta s^2$ between two space-time events $(t_0,x_0)$ and $(t,x)$ is lightlike, that is $\Delta s^2 = (t-t_0)^2 - (x-x_0)^2 = 0$, where we now set the speed of light in vacuum $c=1$ for readability. The space-time event $(t_0,x_0)$ may be identified as the origin in space and time associated with the photon's creation. Equivalently, in $\mathcal H_{\text{GEB}}$, the light-cone constraint in spacetime emerges through the constraint equation $\Box \ket{\psi} = 0$, where we define $\Box = \hat \partial^2_t -\hat \partial^2_x $ the d’Alembertian operator acting on $\mathcal H_{\text{GEB}}$, with $\hat \partial^2_\mu = \int dt dx \; \partial^2_\mu\ket{t,x}\bra{t,x}$; $\mu = t,x$. For a right-propagating photon, $\psi_{\text{GEB}}(t,x)$ then needs to read $\psi_{\text{GEB}}(t,x) = \varphi\left(t-x\right)$, and in turn the GEB state is expressed as
\begin{equation}
\label{Equation GEB state under light-cone constraint}
    \ket{\psi}_{\text{GEB}} = \int_{\mathds R^2}dtdx\; \varphi\left(t-x\right)\ket{t,x}_{\text{GEB}}.
\end{equation}
This means that the photon's worldline lies along a lightlike trajectory $u=t-x$ distributed through the probability amplitude $\varphi\left(u\right)$, such that the specific light cone -- and hence the worldline defined along it -- remains undetermined. This can be mathematically formulated by rewriting the state $\ket{\psi}_{\text{GEB}}$ [Eq.~\eqref{Equation GEB state under light-cone constraint}]
\begin{equation}
\label{Equation light-cone constraint state with light-cone improper}
    \ket{\psi}_{\text{GEB}} = \int_{\mathds R}du\;\varphi(u)\ket{u}_{\text{light-cone}},
\end{equation}
where $\ket{u}_{\text{light-cone}} = \int_{\mathds R} dx\; \ket{x+u,x}$ is an unphysical 2D vector state representing an infinitely spatially delocalized photon along a given light-cone. In other words, while the light-cone constraint $\Delta s^2=0$ restricts the photon to propagate along lightlike worldlines, the probability amplitude $\varphi(u)$ determines how the quantum state is distributed over the family of such worldlines labeled by $u$. As it stands, the states Eq.~\eqref{Equation GEB state under light-cone constraint} and thus Eq.~\eqref{Equation light-cone constraint state with light-cone improper} are ill-defined because we reduce the relevant two-dimensional event Hilbert space represented by $(t,x)$ to a one-dimensional 'light-cone' Hilbert space represented by $u$. In Sup. Mat.~\textcolor{blue}{I}, we provide a straightforward normalization prescription, where $\varphi(u)$ in Eq.~\eqref{Equation light-cone constraint state with light-cone improper} is interpreted as a square-integrable wavefunction defined on the family of light cones labelled by $u$ instead of on the full $(t,x)$ space-time.
\subsection{Space-Time Foliation: Detector as a Reference}
\label{Space-time foliation: detector as a reference}
Incorporating time in quantum theory through a quantum clock~\cite{Quantumclockgiovannetti2015quantum,Quantumclockambiguimarletto2017evolution,Quantumclockmoreva2017quantum,Quantumclocksmith2019quantizing,Quantumclockmaccone2020quantum,Quantumclockfavalli2020time,Quantumcloclfoti2021time,Quantumclockreviewaltaie2022time,Quantumclockkuypers2025measuring} can be inferred by operating a system/quantum clock foliation of the space-time Hilbert space $\mathcal H_{\text{GEB}} \rightarrow \mathcal H_C\otimes \mathcal H_S$, that is $\ket{t,x}_{\text{GEB}} \rightarrow \ket{t}_C\otimes \ket{x}_S$. This amounts to splitting the event into a quantum clock encoding the time coordinate and a system carrying the spatial coordinate, where the clock is to be understood as the time reference from which the system's position in space is defined~\cite{Eventbasedgiovannetti2023geometric}. 
Time evolution then emerges through entanglement between the clock and the system, with so-called \textit{history} states~\cite{Quantumclockgiovannetti2015quantum,Quantumclockmaccone2020quantum}
\begin{equation}
\label{Equation general expression of the history state}
    \ket{\psi}_{\text{clock}} = \int_{\mathds R}dt\; \ket{t}_C\otimes\ket{\psi(t)}_S,
\end{equation}
where
\begin{equation}
\label{Equation state space conditioned on time}
    \ket{\psi(t)}_S = \int dx\;\psi_{\text{PW}}(x|t)\ket{x}_S.
\end{equation}
$\ket{\psi(t)}_S$ is the system state \textit{conditioned} on the clock state being $\ket{t}_C$. That said, we argue that one can equivalently operate the foliation $\mathcal H_{\text{GEB}} \rightarrow \mathcal H_S\otimes \mathcal H_D$ in terms of a \textit{position reference} attached to a detector of Hilbert space $\mathcal H_D$, in which case 
\begin{equation}
\label{Equation space-time foliation with detector}
    \ket{\psi}_{\text{det}} = \int_{\mathds R}dx\;\ket{\psi(x)}_S\otimes\ket{x}_D,
\end{equation}
where 
\begin{equation}
\label{Equation state time conditioned on space}
    \ket{\psi(x)}_S = \int dt\; \psi_{\text{PW}}(t|x)\ket{t}_S.
\end{equation}
Thus, Eq.~\eqref{Equation space-time foliation with detector} realizes a detector-conditioned version of the Page-Wootters mechanism. Previous constructions~\cite{Quantumclockgiovannetti2015quantum,Quantumclockmaccone2020quantum} introduce time as a relational quantum variable through a clock degree of freedom, such that the system observables are conditioned on a clock reading. In our work, the construction is reversed. The detector position provides the reference frame relative to which the system's, and more specifically the photon's, time-of-arrival degree of freedom is defined. The resulting foliation replaces $\ket{\psi(t)}_S$ [Eq.~\eqref{Equation state space conditioned on time}] -- the state conditioned on a \textit{clock} reading -- with $\ket{\psi(x)}_S$ [Eq.~\eqref{Equation state time conditioned on space}] -- the state conditioned on a \textit{detector} reading. This conceptual leap is possible because the GEB framework introduced in Sec.~\ref{Section Geometric event-based quantum mechanics} places space and time on an equal footing, thereby allowing one to select either as the reference frame relative to which the other is defined~\footnote{Rigorously, the correspondence between $\mathcal H_{\text{GEB}}$ and the space-time foliation $\mathcal H_{\text{GEB}}\rightarrow \mathcal H_C\otimes \mathcal H_S$ -- and as such $\mathcal H_{\text{GEB}}\rightarrow \mathcal H_S\otimes \mathcal H_D$ as well -- requires introducing the rigged-extended version  $\mathcal H_{\text{GEB}}^+$ of $\mathcal H_{\text{GEB}}$ accounting for Dirac distributions; see Eq.~(40) of~\cite{Eventbasedgiovannetti2023geometric}.}. The detector-conditioned description [Eq.~\eqref{Equation state time conditioned on space}] is appropriate for TF CV quantum teleportation because the quantity being teleported is the photon's time of arrival, that is the time conditioned on the detector's position. \\ Furthermore, selecting single photons as the carrier of the time degree of freedom is supported by the canonical quantization of~\cite{TFCVfabre2022time}. Concretely, let $\hat a_s(\omega)$ be the bosonic operator associated with the photon's frequency $\omega$ and an auxiliary mode $s$, with bosonic CCRs $[\hat a_s(\omega),\hat a_{s'}^\dagger(\omega')]=\delta_{ss'}\delta(\omega-\omega')$. In the narrow-band approximation, the frequency range can be extended over the entire real line, so that one can define the frequency and time operators $\hat\omega_s=\int d\omega\,\omega\,\hat a_s^\dagger(\omega)\hat a_s(\omega)$ and $\hat t_s=\int dt\,t\,\hat a_s^\dagger(t)\hat a_s(t)$, with $\hat a_s(t)$ the Fourier transform of $\hat a_s(\omega)$. In the single-photon subspace, $\hat t_s$ and $\hat \omega_s$ obey the CCR $[\hat t_s,\hat\omega_s]=i$, as used throughout this work. This extension also sidesteps Pauli's objection~\cite{pauliObjection2012general,Quantumclockleon2017pauli}, which forbids a self-adjoint time operator conjugate to an energy operator whose spectrum is bounded from below. For a photon whose spectrum lies far from $\omega=0$ relative to its bandwidth, the positivity constraint is inactive up to negligible corrections, so the CCR holds as an approximation rather than as an exact statement. Together with the detector-conditioned foliation of Eq.~\eqref{Equation space-time foliation with detector}, this constitutes the definition of the time of arrival that we adopt.\\
In what follows, we omit the extra $_\text{det}$ labels.
\subsection{Fixed Detector}
The space-time foliation in Eq.~\eqref{Equation space-time foliation with detector} shows that the detector's position is imprinted on the definition of the photon's time-of-arrival degree of freedom, which indicates that one cannot teleport the latter without the former. In practice, however, the detector's position $x$ may be reduced to a classical label $x_D$ by fixing the detector's position, that is 
\begin{equation}
    \ket{\psi(x)}_S\ket{x}_D\rightarrow \ket{\psi(x_D)}_S
\end{equation}
in Eq.~\eqref{Equation space-time foliation with detector}. More specifically for right-propagating photons, Eq.~\eqref{Equation GEB state under light-cone constraint}, the space-time foliation conditioned on a given detector position $x=x_D$ leads to
\begin{equation}
\label{Equation photon state with space-time foliation and fixed detector}
    \ket{\psi} = \int_{\mathds R}dt\;\varphi(t-x_D)\ket{t}_S.
\end{equation}
In our work, the detector reference frame is explicitly included through the very definition of the time-of-arrival degree of freedom, defined conditionally on the detector's position. To the question "What information is to be teleported?" for quantum teleportation of the photon's TF spectrum the answer is thus the raw wavefunction $\varphi(t)$ \textit{and} the detector's position $x_D$ with respect to which the time of arrival is defined.  
\section{Quantum teleportation protocol}
\label{Section Quantum teleportation protocol}
We now present the single-photon TF CV quantum teleportation protocol, adapting the proposal of Braunstein and Kimble~\cite{CVQTeleportationbraunstein1998teleportation,CVQTeleportationpirandola_quantum_2006} considered for quadrature CVs.
\subsection{Single- and Two-Photon States}
\label{Section protocol Single- and two-photon states}
From Eq.~\eqref{Equation photon state with space-time foliation and fixed detector}, a single-photon state simply reads
\begin{equation}
\label{Equation spatial foliation for a fixed detector position 2}
    \ket{\psi}_{\text{in}} = \int_{\mathds R}dt\;\varphi(t)\ket{t+x_{\text{in}}}_{\text{in}},
\end{equation}
where we shifted the integral $t\rightarrow t+x_{\text{in}}$. The light-cone constraint hence trivially imprints the detector reference frame on the time degree of freedom. Fixing the detector's position in the space-time foliation does not imply, however, that all parties in the quantum teleportation protocol share the same detector reference position. \\
Similarly to $\ket{\psi}_{\text{in}}$ [Eq.~\eqref{Equation spatial foliation for a fixed detector position 2}], we express the two-photon quantum channel shared by the two parties Alice ($A$) and Bob ($B$) with the spatial foliation of Eq.~\eqref{Equation space-time foliation with detector} as 
\begin{equation}
\label{Equation two-photon state with space-time foliation with detectors}
    \ket{\phi}_{AB} =\int_{\mathds R^2}dt_Adt_B\; \phi(t_A,t_B)
    \ket{t_A+x_A,t_B+x_B}_{AB},
\end{equation}
where as previously mentioned, the detectors' positions $x_A$ and $x_B$ conditioning time are not necessarily identical. The initial tripartite state then reads $ \ket{\psi}_{\text{in},AB} = \ket{\psi}_{\text{in}} \otimes \ket{\phi}_{AB}$, where $\ket{\psi}_{\text{in}}$ is the unknown state Alice wishes to teleport to Bob and $\ket{\phi}_{AB}$ the so-called quantum channel. 
\subsection{Protocol Steps}
\label{Section protocol protocol steps}
\begin{figure*}
    \centering
    \includegraphics[width=0.9\linewidth]{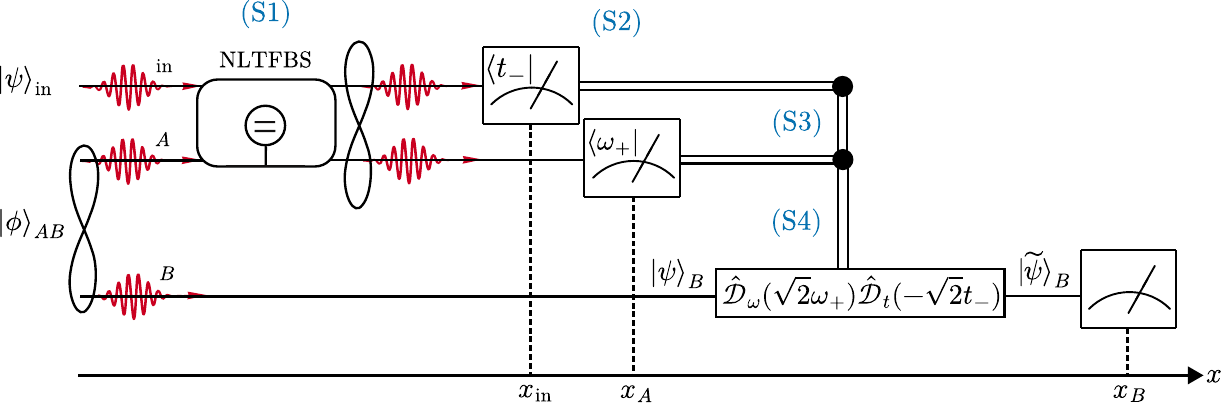}
    \caption{Single-photon TF CV quantum teleportation protocol scheme. Step (S1) corresponds to the TF mode mixing realized by a nonlinear TFBS, here abbreviated as NLTFBS. Step (S2) consists of TF measurements which are assumed to be projective. (S1) and (S2) together constitute the Bell measurement stage, which we define as the measurement of the commuting collective operators $\hat t_-$ and $\hat \omega_+$. In step (S3), Alice sends the measurement outcome obtained in (S2) to Bob through the classical channel. Bob then performs TF displacements to his photon in step (S4). Apart from labeling the different parties, $\text{in}$, $A$ and $B$ indicate different auxiliary degrees of freedom, for instance different rails along which the photons propagate. $x_{\text{in}}$, $x_A$ and $x_B$ denote the positions of the detectors where the photons are to be measured.}
    \label{Figure protcol scheme}
\end{figure*}
\subsubsection{Nonlinear Time-Frequency Beamsplitter}
(S1) First, Alice is to perform a mode-mixing entanglement operation $\ket{t,t'}_{\text{in},A} \rightarrow \ket{\frac{t'-t}{\sqrt{2}},\frac{t'+t}{\sqrt{2}}}_{\text{in},A}$, with the target and Alice's photons now time entangled, see Fig.~\ref{Figure protcol scheme}. This entanglement operation can equivalently be understood as the transformation into the collective operators $\hat t_\pm = \left(\hat t_{\text{A}}\pm\hat t_{\text{in}}\right)/\sqrt{2}$ and $\hat \omega_\pm = \left(\hat \omega_\text{A}\pm\hat \omega_{\text{in}}\right)/\sqrt{2}$ of the time and frequency operators $\hat t_s$ and $\hat \omega_s$ (see Sec.~\ref{Space-time foliation: detector as a reference}), restricted to the single-photon subspace, with $s = A, \text{in}$. Implementing this beamsplitter-equivalent mode mixer requires nonlinear optical interactions, and more specifically a quartic Hamiltonian $\hat H_{\text{TFBS}}\sim \iint dtdt'\;\hat a_{\text{A}}^\dagger\left(\frac{t+t'}{\sqrt{2}}\right)\hat a_{\text{in}}^\dagger\left(\frac{t-t'}{\sqrt{2}}\right)\hat a_{\text{A}}(t)\hat a_{\text{in}}(t') + \text{h.c.}$, with $\hat a_{\text{A}}(t)$ and $\hat a_{\text{in}}(t')$ the bosonic operators of modes $A$ and $\text{in}$. The beamsplitter-equivalent mode mixer is hence referred to as nonlinear TF beamsplitter (TFBS). The nonlinear TFBS enables a complete Bell-state measurement and, in principle, allows single-photon TF CV quantum teleportation to achieve unit efficiency, provided that perfect mode mixing can be realized. In practice, however, achieving efficient photon-photon quartic interactions remains challenging. Recent works~\cite{alushi2023waveguide,FBSmeguebel2025generation} have shown that this limitation can be circumvented by using matter as an intermediary, whereby a quadratic Hamiltonian can reproduce the entangling operation of the nonlinear TFBS.
\subsubsection{Time-Frequency Measurements}
(S2) Next, Alice performs a TF measurement in ports in and $A$ on her state and the target state and records the outcome $(t_-,\omega_+)$, which amounts to the $\bra{t_-}_\text{in}\bra{\omega_+}_A$ projective measurement, see Fig.~\ref{Figure protcol scheme}. Alice can measure both $\hat t_-$ and $\hat \omega_+$ simultaneously because, from the CCR~\cite{TFCVfabre2022time}, the collective operators $\hat t_-$ and $\hat \omega_+$ commute. The time of arrival can be measured by time-resolved single-photon detection. The frequency can be measured either with a diffraction-grating spectrometer mapping frequency onto position before a single-photon array~\cite{jirsa2025fast}, or by frequency-to-time mapping~\cite{avenhaus2009fiber,davis2017pulsed}, in which a dispersive element maps frequency onto arrival time followed by time-resolved detection. The projection $\bra{t_-}_\text{in}$ collapses $t_A$ to $t_A = t+\sqrt{2}t_-+x_{\text{in}}-x_A$. The term $x_{\text{in}}-x_A$ results from our relational treatment of time, defined conditionally on the detector's position. This term is absent in the standard Braunstein-Kimble formulation because the reference frame with respect to which the states and their quadratures are defined is taken implicitly as part of the background, rather than being represented as an additional degree of freedom. After the TF projective measurement, the tripartite state then collapses, up to global phases, to Bob's single-photon state as 
\begin{equation}
\begin{split}
    \ket{\psi}_B = &\int_{\mathds R}dt\;\varphi(t)e^{-it\sqrt{2}\omega_+}\\
    &\cross \int_{\mathds R}dt_B\phi\left(t+\sqrt{2}t_-+x_{\text{in}}-x_A,t_B\right)\ket{t_B+x_B}_B,
\end{split}
\end{equation}
where we used the Fourier-transform relation between the time and frequency bases $\ket{\omega} = \frac{1}{\sqrt{2\pi}}\int_{\mathds R}dt\; e^{i\omega t}\ket{t}$. Because of the Bell measurement stages (S1) and (S2), the quantum channel has inherited a time shift and a frequency shift that Bob tries to compensate in (S4).
\subsubsection{Classical Communications and Conditional Displacements}
(S3) The recorded data $(t_-,\omega_+)$ is then sent to Bob through a classical communication channel. In step (S4), Bob, as shown in Fig.~\ref{Figure protcol scheme}, performs a conditional correction to his state with associated displacement operators~\cite{fabre2020generation,TFCVfabre2022time} $\hat{\mathcal D}_\omega(\sqrt{2}\omega_+)\hat{\mathcal D}_t(-\sqrt{2}t_-)$, which produces, up to a global phase, the following state 
\begin{equation}
\label{Equation output state at Bob's stage after corrections}
\begin{split}
    \ket{\psi}_B\rightarrow &\int_{\mathds R}\int_{\mathds R}dt dt_B\;\varphi(t)e^{-i\sqrt{2}\omega_+(t-t_B)} \\
    &\cross \phi(t+\sqrt{2}t_-+x_{\text{in}}-x_A,t_B)\ket{t_B+x_B-\sqrt{2}t_-}_B,
\end{split}
\end{equation}
where $\hat{\mathcal D}_{t}(\tau)\ket{t} = \ket{t+\tau}$ and $\hat{\mathcal D}_\omega(\nu)\ket{t} = e^{i\nu t}\ket{t}$ for time $\tau$ and frequency shifts $\nu$, respectively. Frequency shifts in particular can be performed with electro-optic modulators (EOMs) or acousto-optic modulators (AOMs). \\
The state Eq.~\eqref{Equation output state at Bob's stage after corrections} is a distorted version of the input wavepacket Eq.~\eqref{Equation spatial foliation for a fixed detector position 2}, resulting from  a projection of Bob's state onto the target state weighted by the wavefunction $\phi(t_A,t_B)$ associated with the quantum channel. \\ \\
In Appendix~\ref{Appendix Protocol calculations in chronocyclic phase space}, we derive the expressions of Bob's output state [Eq.~\eqref{Appendix Equation general average output state}] for general input states
and quantum channels. The calculations therein are performed in the single-photon TF phase space, manipulating the chronocyclic Wigner function~\cite{fabre2020generation,TFCVfabre2022time}.
\subsection{Input-Output State Fidelity} 
\label{Section Input-output state fidelity}
\subsubsection{Ideal Quantum Channel}
For an ideal quantum channel, the quantum state shared by Alice and Bob follows an EPR-state distribution, that is to say $\phi(t_A,t_B) = \delta(t_A-t_B)$. In this case, the state Eq.~\eqref{Equation output state at Bob's stage after corrections} reduces to 
\begin{equation}
    \ket{\psi}_{B,\text{ideal}} = \int_{\mathds R}dt\;\varphi(t)\ket{t+x_{\text{in}}+\Delta x_{BA}}_B,
\end{equation}
where $\Delta x_{BA} = x_B-x_A$ is the detector-position mismatch, which we will extensively discuss shortly. In other words, the teleported state at Bob's side is a displaced version of the target state $\ket{\psi}_{\text{in}}$ [Eq.~\eqref{Equation spatial foliation for a fixed detector position 2}], where both the raw wavefunction $\varphi(t)$ \textit{and} the reference position $x_{\text{in}}$ have been teleported. For $\Delta x_{BA} = 0$, Bob directly retrieves the time-of-arrival degree of freedom $\varphi(t)$ with respect to a detector placed at $x=x_{\text{in}}$, even though his own detector need not be at this position, that is $x_B\neq x_{\text{in}}$ in general: it appears as if Bob's detector has been translated by $x_{\text{in}}-x_B$ after the teleportation. \\
The input-output state fidelity $\mathcal F = |\;_\text{in}\bra{\psi}\ket{\psi}_{B,\text{ideal}}\;|^2$ is equal to the temporal  overlap of the photon’s wavepacket at retarded times
\begin{equation}
\label{Equation teleportation fidelity for ideal channels and measurements}
    \mathcal F = \left|\int_{\mathds R}dt\; \varphi(t)\varphi^*\left(t+\Delta x_{BA}\right)\right|^2 \leq 1.
\end{equation}
The fidelity is therefore a direct measure of the autocorrelation function of the target wavefunction $\varphi(t)$ at lag $\tau=\Delta x_{BA}$ ($c=1$). The detector-position mismatch $\Delta x_{BA}$ reduces the fidelity when it is left uncorrected, even assuming ideal quantum and classical communication channels and ideal measurements. Importantly, this does not imply that Alice's and Bob's detectors must be located at the same position. Rather, $\Delta x_{BA}$ should be understood as the residual mismatch between Alice's and Bob's detector-position references that enters the teleportation correction. Alice could communicate her detector position $x_A$ to Bob, but this alone is not sufficient unless Alice and Bob have first established a common position reference. We therefore use the term \textit{synchronization} to denote the prior establishment of such a common position reference between Alice and Bob. Without this synchronization, the coordinate values assigned by Alice and Bob may refer to different origins, so that even if Alice communicates $x_A$ to Bob, the relative detector-position mismatch entering the teleportation correction cannot in general be determined. If, however, Bob knows his own position $x_B$ with respect to the same position reference as Alice's $x_A$, then $\Delta x_{BA}=x_B-x_A$ can be determined and the corresponding temporal displacement exactly corrected, allowing, under the present ideal assumptions, the fidelity to reach unity. \\
Interestingly, this synchronization need not be established independently of the teleportation protocol: the protocol can calibrate its own detector-position reference. For a known input state, evaluating the overlap between the teleported output and the target states enables Bob to vary the correction needed to cancel $\Delta x_{BA}$ and maximize the output fidelity, thereby identifying the correction that compensates the detector-position reference mismatch. In this way, Alice and Bob can achieve \textit{remote synchronization} of their detector positions directly through the teleportation protocol itself, without communicating their individual detector coordinates -- which would require Alice to reveal her detector position -- and without agreeing beforehand on a common position reference. \\
A variety of methods have been proposed and carried out to measure the spectro-temporal properties at the single-photon level, therefore allowing one to assess the overlap between the teleported output state and the target state in our protocol. These include Hong-Ou-Mandel interferometry~\cite{wasilewski2007spectral} and spectrally-resolved Hong-Ou-Mandel interferometry~\cite{thiel2020single} with a reference state, and self-referencing methods based on electro-optic shearing~\cite{davis2018measuring}.
\subsubsection{Input Gaussian Wavepacket -- Fidelity-Bandwidth Trade-Off}
Let us now take for example a Gaussian input wavepacket~\cite{sakurai2020modern} $\varphi(t) = \left(2\pi\sigma_t^2\right)^{-1/4}\;e^{-\frac{(t-t_0)^2}{4\sigma_t^2}}e^{i\omega_0(t-t_0)}$. The fidelity Eq.~\eqref{Equation teleportation fidelity for ideal channels and measurements} is immediately computed as 
\begin{equation}
\label{Equation fidelity for a Gaussian wavepacket with an ideal quantum channel}
    \mathcal F_{\text{Gaussian}} = e^{-\frac{\left(\Delta x_{BA}\right)^2}{4\sigma_t^2}}.
\end{equation}
For a Gaussian single-photon input, the ideal teleportation fidelity decreases smoothly, as a Gaussian, with increasing $|\Delta x_{BA}|$, such that larger detector-position mismatch results in lower teleportation fidelity. This entails that the relevant scale is set by the temporal width $\sigma_t$. The narrower the photon in time, the smaller the allowable spatial mismatch before the fidelity drops noticeably. In practice, adopting the $\mathcal F =1/2$ threshold, which we prove in Sec.~\ref{Section Fidelity threshold benchmarks} to be the classical limit threshold for single-photon Gaussian wavepackets, the fidelity falls below this value for $|\Delta x_{BA}| > \Delta x_{BA}^{(1/2)} = 2c\sigma_t\sqrt{\ln2}$, \textit{i.e.} about $5~\mu$m for $\sigma_t = 10$ fs, $50~\mu$m for $\sigma_t = 100$ fs and $0.5$ mm for $\sigma_t = 1$ ps, and up to millimeters or centimeters for broader wavepackets. Our result therefore provides a concrete and experimentally meaningful bandwidth/synchronization trade-off, as it shows that higher temporal localization makes the teleportation protocol increasingly sensitive to imperfect detector-position reference-frame synchronization. \\ \\
The Gaussian wavepacket contains a single temporal scale given by its standard deviation. States with internal temporal structure, however, provide additional relational time references, which modify the detector-dependent teleportation fidelity structure. As we shall see next for two kinds of non-Gaussian states, this can give rise to nontrivial fidelity behaviors, including revival phenomena, which can be exploited for even more accurate detector-position synchronization. 
\subsubsection{Fidelity Revivals}
\paragraph{Time-Frequency Cat States} Consider first an input wavepacket composed of a sum of two Gaussians $\varphi_{\text{cat}}(t) = \mathcal N_{\text{cat}} \left( g(t-t_0)e^{i\omega_0(t-t_0)} 
+ e^{i\theta}g(t+t_0)e^{-i\omega_0(t+t_0)}\right)$, where $\mathcal N_{\text{cat}} = \left(2\left(1+\cos\theta e^{-2|\alpha|^2}\right)\right)^{-1/2}$ is a normalization constant, $g(t)=\left(2\pi s_t^2\right)^{-1/4}\;e^{-\frac{t^2}{4s_t^2}}$ a Gaussian distribution and $\theta$ a relative phase. We call the input state of wavefunction $\varphi_{\text{cat}}(t)$ a TF \textit{cat state} by analogy with quadrature cat states~\cite{ourjoumtsev2007generation,albert_performance_2018}, with in particular the equivalent coherent-state squared absolute value $|\alpha|^2 = \frac{t_0^2}{4s_t^2}+s_t^2\omega_0^2$. Note that the central frequency of single-photon wavepackets should be positive, while here one of the wavepackets has $-\omega_0$ central frequency. We circumvent this ambiguity by defining the central frequencies with respect to an arbitrary frequency $\omega_{\text{ref}}$, which are then understood as detunings with respect to $\omega_{\text{ref}}$.  
\begin{figure*}
    \centering
    \includegraphics[width=\linewidth]{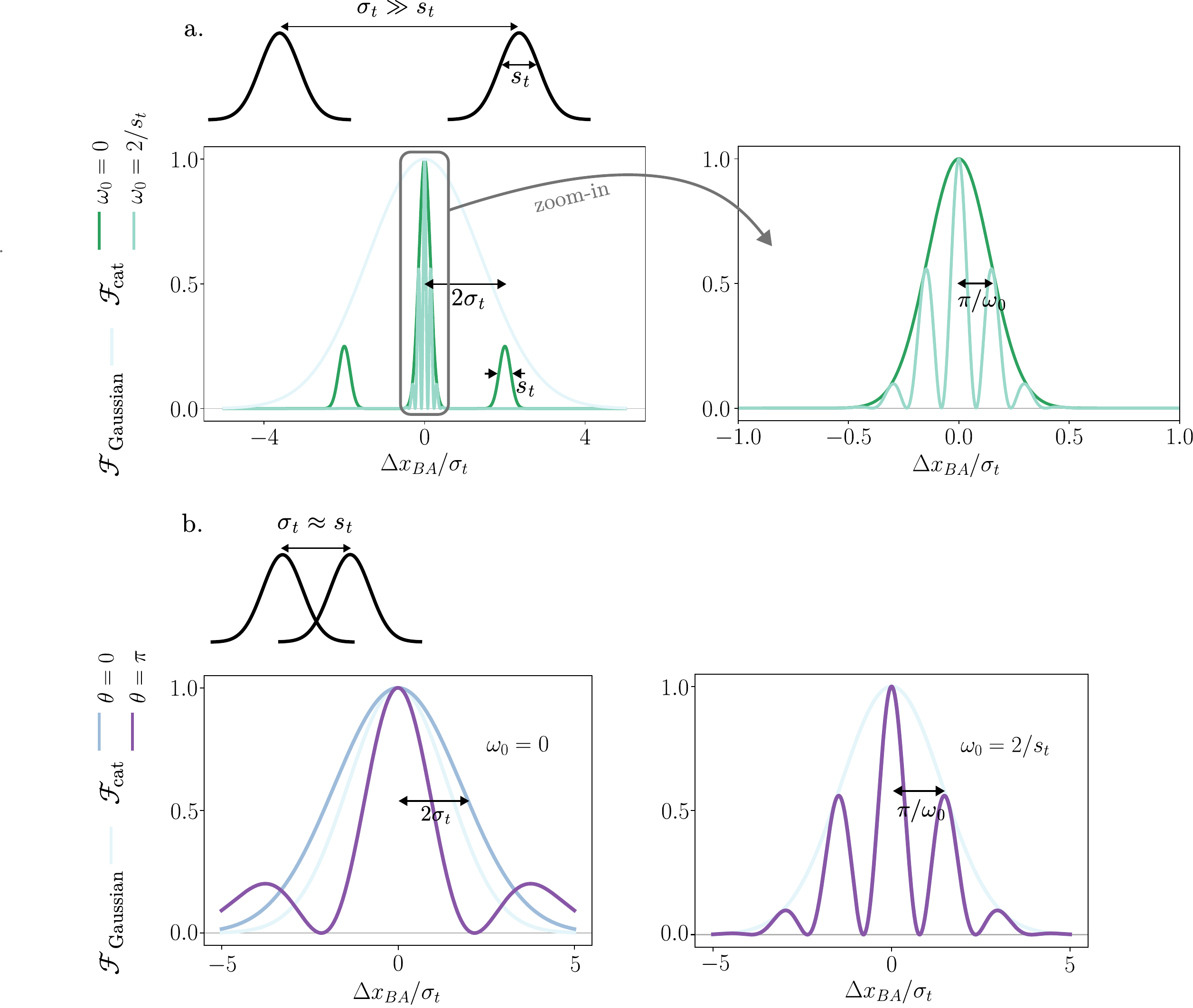}
    \caption{Fidelities $\mathcal F_\text{Gaussian}$ [Eq.~\eqref{Equation fidelity for a Gaussian wavepacket with an ideal quantum channel}] (\raisebox{0.5ex}{\textcolor{e5f5f9}{\rule{0.5cm}{1.5pt}}}) and $\mathcal F_{\text{cat}}$ [Eq.~\eqref{Equation TF cat state fidelity for ideal quantum channel}] for $\sigma_t = t_0 \gg s_t$ (panel a.) and $\sigma_t \approx s_t$ (panel b.), where $\sigma_t$ is the single Gaussian wavepacket standard deviation, $2t_0$ the time distance between the two cat state Gaussian wavepackets of standard deviation $s_t$. In each panel, the black curves schematically represent the input state $\varphi(t)$ under consideration. In the $\sigma_t \gg s_t$ regime, we plot the fidelity $\mathcal F_{\text{cat}}$ for $\omega_0 = 0$ (\raisebox{0.5ex}{\textcolor{2ca25f}{\rule{0.5cm}{1.5pt}}}) and for $\omega_0 = 2/s_t$ (\raisebox{0.5ex}{\textcolor{99d8c9}{\rule{0.5cm}{1.5pt}}}). Although the full width at half maximum (FWHM) characterizing the three main peaks for $\omega_0=0$ is  $\sqrt{2}s_t\times2\sqrt{2\ln(2)}$, we simply indicate the time scale $s_t$ for readability. For the $\sigma_t \approx s_t$ regime, we plot the fidelity $\mathcal F_{\text{cat}}$ for relative phases $\theta = 0$ (\raisebox{0.5ex}{\textcolor{9ebcda}{\rule{0.5cm}{1.5pt}}}) and $\theta=\pi$ (\raisebox{0.5ex}{\textcolor{8856a7}{\rule{0.5cm}{1.5pt}}}) and for $\omega_0=0$ and $\omega_0=2/s_t$. The parameters $s_t$ and $\omega_0$ are normalized by $\sigma_t$. The fidelities are all given with respect to $\Delta x_{BA}/\sigma_t$, the detector mismatch normalized by $\sigma_t$.}
    \label{Figure TF cat state fidelity revivals}
\end{figure*}
For an ideal quantum channel, the fidelity Eq.~\eqref{Equation teleportation fidelity for ideal channels and measurements} can be computed for $\varphi_{\text{cat}}$ as
\begin{widetext}
\begin{equation}
\label{Equation TF cat state fidelity for ideal quantum channel}
    \mathcal F_{\text{cat}} = \frac{\left|2\cos(\omega_0\Delta x_{BA}) e^{-\frac{\Delta x_{BA}^2}{8s_t^2}}+e^{-i\theta} e^{-\frac{(\Delta x_{BA}+2t_0)^2}{8s_t^2}}e^{-2\omega_0^2s_t^2}+e^{i\theta}e^{-\frac{(\Delta x_{BA}-2t_0)^2}{8s_t^2}}e^{-2\omega_0^2s_t^2}\right|^2}{4\left(1+\cos\theta e^{-2|\alpha|^2}\right)^2}.
\end{equation}
\end{widetext}
In the $s_t \ll t_0$ regime, that is to say for well-separated lobes, $\mathcal F_{\text{cat}}$ simplifies to 
\begin{equation}
\begin{split}
\label{Equation TF cat state fidelity for ideal quantum channel well separated}
    \mathcal F_{\text{cat}} &\approx \cos^2(\omega_0\Delta x_{BA})e^{-\frac{\Delta x_{BA}^2}{4s_t^2}}\\
    &+\frac{1}{4}e^{-\frac{(\Delta x_{BA}+2t_0)^2}{4s_t^2}}e^{-4\omega_0^2s_t^2}+\frac{1}{4}e^{-\frac{(\Delta x_{BA}-2t_0)^2}{4s_t^2}}e^{-4\omega_0^2s_t^2}.
\end{split}
\end{equation}
In other words, in this regime, the fidelity $\mathcal F_{\text{cat}}$ no longer depends on the relative phase $\theta$. Computation details are provided in Sup. Mat.~\textcolor{blue}{II}. The fidelity $\mathcal F_{\text{cat}}$ [Eq.~\eqref{Equation TF cat state fidelity for ideal quantum channel well separated}] does not, unlike the fidelity $\mathcal F_{\text{Gaussian}}$ [Eq.~\eqref{Equation fidelity for a Gaussian wavepacket with an ideal quantum channel}] obtained for a single Gaussian wavepacket, simply plummet to zero from its maximum $\mathcal F_{\text{cat}} = 1$ at $\Delta x_{BA} = 0$. Rather, the fidelity exhibits two partial revivals to $\mathcal F_{\text{cat}} \approx \frac{1}{4}e^{-4\omega_0^2s_t^2}$ at $\Delta x_{BA} = \pm 2t_0$. This result stands to reason inasmuch as a $+2t_0$ (-$2t_0$) detector mismatch corresponds to situations where the output state early (late) lobe overlaps with the target state late (early) lobe. \\
A lobe in the TF cat state acts as a time reference for the other, giving meaning to the denomination 'late' and 'early' lobes. In this respect, the fidelity revivals at $\Delta x_{BA}= \pm 2t_0$ may further be interpreted as the \textit{internal} reference of the TF cat state impinging on the \textit{external} reference represented by the detectors' positions: $\Delta x_{BA} \rightarrow \Delta x_{BA} \pm 2t_0$.  \\
The fidelity decay from the central fidelity peak at $\Delta x_{BA} = 0$ shows an oscillatory behavior of frequency $\omega_0$. These oscillations amount to additional periodic partial fidelity revivals at $\Delta x_{BA} = \pi k/\omega_0$, $k\in \mathds Z$ with a fidelity $\mathcal F_{\text{cat}} = e^{-\frac{\pi^2k^2}{4\omega_0^2s_t^2}}$. These revivals can be observed for $\omega_0$ large compared with $1/s_t$, as the decreasing Gaussian would otherwise suppress them.  Consequently, while for $\omega_0\gg 1/s_t$ the side revivals at $\Delta x_{BA} = \pm 2t_0$ are suppressed and the revivals from the central peak oscillations enhanced, it is the other way around for $\omega_0 \ll 1/s_t$, see Fig.~\ref{Figure TF cat state fidelity revivals} (panel a.) for $\omega_0=2/s_t$ and $\omega_0=0$, respectively. \\ \\
In the $s_t\approx t_0$ regime with $\mathcal F_{\text{cat}}$ [Eq.~\eqref{Equation TF cat state fidelity for ideal quantum channel}], the $\Delta x_{BA} = \pm 2t_0$ fidelity revivals are suppressed. In particular, for $\omega_0=0$, the TF cat state with $\theta=\pi$ shows destructive interference, taking $t_0=s_t$, for $\cosh(\frac{\Delta x_{BA}}{2s_t}) = e^{1/2}$, that is $\Delta x_{BA} \approx 2s_t$, as can be observed in Fig.~\ref{Figure TF cat state fidelity revivals} (panel b.). For $\omega_0=2/s_t$, only the oscillating central peak at $\Delta x_{BA} = 0$ survives, and because $s_t\approx t_0$, the corresponding fidelity revivals spread over the temporal extent $2t_0$, see Fig.~\ref{Figure TF cat state fidelity revivals} (panel b.).  
\paragraph{Time-Frequency Gottesman-Kitaev-Preskill States} We now take a TF GKP single-photon state~\cite{fabre2020generation,yamazaki_linear_2023,TFGKPpousset2026time} associated with the odd and even components of a frequency comb, analogous to quadrature GKP states~\cite{gottesman2000encoding,brady2024advances}. More precisely, the corresponding even and odd wavefunctions in the time domain are, respectively, $\varphi_{\text{even}}(t) = \mathcal N_{\text{\text{even}}}e^{-\frac{t^2}{4\sigma_t^2}}\sum_{n\in \mathds Z}g(t-2n\overline \tau)$ and $\varphi_{\text{\text{odd}}}(t) = \mathcal N_{\text{odd}}e^{-\frac{t^2}{4\sigma_t^2}}\sum_{n\in \mathds Z}g(t-(2n+1)\overline \tau)$, where $g(t)  = \left(2\pi s_t^2 \right)^{-1/4}e^{-\frac{t^2}{4s_t^2}}$ and $\mathcal N_{\text{even,odd}}$ are normalization constants. Here, $\sigma_t$ denotes the comb envelope standard deviation, $s_t$ the individual peaks standard deviation and $\overline \tau$ the comb periodicity. 
\begin{figure}
    \centering
    \includegraphics[width=\linewidth]{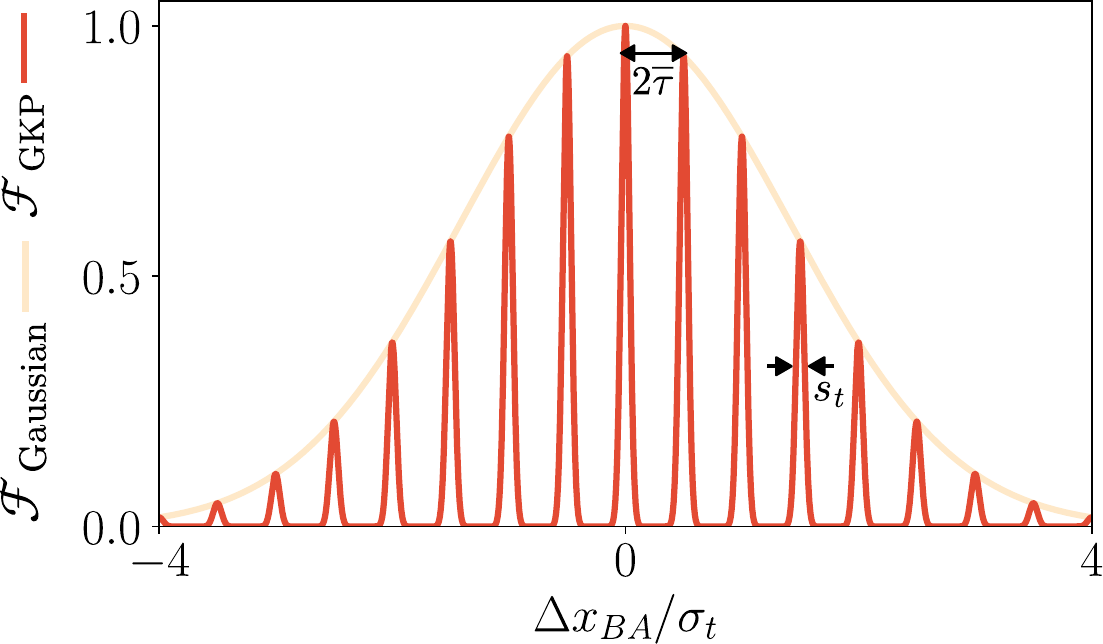}
    \caption{Fidelities $\mathcal F_\text{Gaussian}$ [Eq.~\eqref{Equation fidelity for a Gaussian wavepacket with an ideal quantum channel}] (\raisebox{0.5ex}{\textcolor{fee8c8}{\rule{0.5cm}{1.5pt}}}) and $\mathcal F_{\text{GKP}}$ [Eq.~\eqref{Equation Fidelity for time GKP state wavepacket}]  (\raisebox{0.5ex}{\textcolor{e34a33}{\rule{0.5cm}{1.5pt}}}) for $\overline \tau = \sigma_t/4$ and $s_t = \overline \tau/10$. Although the FWHM characterizing the peaks is $\sqrt{2}s_t\times2\sqrt{2\ln(2)}$, we simply indicate the time scale $s_t$ for readability. The parameters $\overline \tau$ and $s_t$ are normalized by $\sigma_t$. The fidelities are given with respect to $\Delta x_{BA}/\sigma_t$, the detector mismatch normalized by $\sigma_t$.}
    \label{Figure TF GKP state fidelity revivals}
\end{figure}
TF GKP states are promising qubit states, inasmuch as they are constructed to be robust against small temporal and frequency shifts, which can subsequently be corrected~\cite{fabre_teleportation-based_2023}. \\
In the $\sigma_t \gg \overline \tau \gg s_t$ limit, the input-output fidelity reads 
\begin{equation}
\label{Equation Fidelity for time GKP state wavepacket}
   \mathcal F_{\text{GKP}} = \left|\sum_{l\in \mathds Z}e^{-\frac{\overline \tau^2l^2}{2\sigma_t^2}} e^{-\frac{\left(\Delta x_{BA}-2\overline \tau l\right)^2}{8s_t^2}}\right|^2,
\end{equation}
for both the even and odd TF GKP states, where calculation details are given in Sup. Mat.~\textcolor{blue}{III}. As showcased in Fig.~\ref{Figure TF GKP state fidelity revivals}, the fidelity $\mathcal F_{\text{GKP}}$ [Eq.~\eqref{Equation Fidelity for time GKP state wavepacket}] displays partial fidelity revivals of heights approximately equal to $e^{-\frac{\overline \tau^2l^2}{\sigma_t^2}}$, $l\in \mathds Z$, which rise periodically for $\Delta x_{BA} = 2\overline \tau l$ and extend over the $s_t$ time scale. In this light, the teleportation fidelity reproduces the internal temporal lattice carried by the TF states. \\ \\
Taken together, these results show that, for a fixed overall temporal extent, structured TF states can provide a much finer probe of detector-position mismatch than a single Gaussian wavepacket, combining the capture range of a broad envelope with the resolution of narrow peaks. Indeed, when comparing TF cat and GKP states to a single Gaussian wavepacket with the same temporal extent, their fidelity revivals are characterized by widths set by the individual-peak scale $s_t$, which can be much smaller than the total temporal extent. For a TF cat state, this narrow fidelity structure is set by the individual lobe width $s_t$ or by $\pi/\omega_0$, with $\omega_0$ defining the TF cat state. For a TF GKP state, the intrinsic temporal lattice produces a series of periodic revivals separated by $2\overline\tau$, with local precision set by $s_t$. The latter therefore acts as a ruler for detector-position mismatch, with $2\overline\tau$-spaced markers over the full temporal extent and a local resolution determined by $s_t$. These intrinsic temporal structures characterizing the TF cat and GKP states are directly imprinted onto the teleportation fidelity as a function of $\Delta x_{BA}$ and can thus serve as internal references for estimating and correcting the relative detector position. This suggests a two-stage operational use of the protocol, where a known structured TF state is first used to remotely synchronize and calibrate Alice's and Bob's detector-position references by optimizing the teleportation correction, after which an unknown single-photon TF state, such as a Gaussian wavepacket, can be teleported without the fidelity penalty associated with an uncorrected detector-position mismatch.\\ \\
The remaining question is then whether the resulting teleportation fidelity genuinely exceeds what can be achieved without shared quantum resources, which motivates the introduction of appropriate fidelity benchmarks in the next section. 
\subsection{Fidelity Threshold Benchmarks}
\label{Section Fidelity threshold benchmarks}
Assessing whether the proposed protocol exploits genuine quantum resources requires establishing a benchmark for the teleportation fidelity. Exceeding this benchmark demonstrates a quantum advantage over any strategy that relies only on classical resources. For instance, in the absence of shared entanglement, the so-called \textit{measure-and-prepare} classical strategy consists in Alice performing a measurement of her unknown input state $\ket{\psi}_{\text{in}}$, communicating the result to Bob through the classical channel and having Bob prepare an output state conditioned on the measurement result. For a given ensemble $\mathcal A$ of input states, the corresponding classical fidelity threshold (CFT) is defined as the maximum average fidelity $\max\mathcal F_{\text{class}}$ achievable by any measure-and-prepare strategy~\cite{chiribella2014quantum}. \\
In quadrature-based CV protocols, coherent states parameterized by their complex amplitude $\alpha \in \mathds C$ constitute a common choice of input-state  ensemble, for which the CFT is $\max\mathcal F_{\text{class}} = 1/2$~\cite{braunstein2000criteria,hammerer2005quantum}. This threshold is a particular case of a more general 1-to-$M$ cloning threshold~\cite{grosshans2001quantum,cerf2005non}, where the limit $M\rightarrow +\infty$ of infinitely many copies amounts to the measure-and-prepare strategy. The $M=2$ case is associated with a more stringent $\max \mathcal F_{1\rightarrow 2} = 2/3$ threshold for a Gaussian cloner, ensuring that Bob's state is the best possible remaining copy of the input state. This threshold thus provides a criterion for detecting eavesdropping in quantum communication through excess-noise measurements. \\
To the best of our knowledge, no analogous thresholds have been calculated for CV encodings based on single-photon degrees of freedom, in particular for the single-photon TF CV encoding. Since single-photon TF CVs constitute a distinct encoding from optical quadratures, these thresholds cannot simply be assumed and must instead be derived within the present framework.  We now compute the measure-and-prepare and best-copy thresholds for pure single-photon states with a Gaussian wavepacket.
\subsubsection{Measure-and-Prepare Strategy Threshold}
\label{Section Measure-and-Prepare Strategy Threshold}
Consider an input ensemble $\mathcal A = \{ \ket{t_0,\omega_0;x_{\text{in}}}\}$, where $\ket{t_0,\omega_0;x_{\text{in}}}$ is a pure single-photon state $ \ket{t_0,\omega_0;x_{\text{in}}} = \int_{\mathds R}dt\;\varphi_{t_0,\omega_0}(t-x_{\text{in}})\ket{t}$, with a Gaussian wavepacket $\varphi_{t_0,\omega_0}(t)= \left(2\pi\sigma_t^2\right)^{-1/4}\;e^{-\frac{(t-t_0)^2}{4\sigma_t^2}}e^{i\omega_0(t-t_0)}$. The $(t_0,\omega_0)$ parameters are distributed according to a Gaussian distribution $ p(t_0,\omega_0)=\frac{\lambda}{2\pi}e^{-\lambda\left(\frac{t_0^2}{4\sigma_t^2}+\sigma_t^2\omega_0^2\right)}$ with the inverse variance $\lambda$ and with a fixed standard deviation $\sigma_t$ for the whole $\mathcal A$ ensemble. As for TF cat states, we define the central frequency with respect to an arbitrary frequency $\omega_{\text{ref}}$, that is in terms of a frequency detuning, so as to extend the frequency integrals over the full real line. \\
The derivation of the measure-and-prepare strategy fidelity threshold as proposed in~\cite{braunstein2000criteria} -- with optimality over \textit{all} measure-and-prepare strategies  proven in~\cite{hammerer2005quantum} --  manipulates coherent states, which are physically very different from single-photon Gaussian wavepackets. The two are nonetheless related by a direct correspondence, which we now introduce. \\
Consider $\text{TF}^{(x)}_{\alpha}: \ket{t,\omega;x}\rightarrow \ket{\alpha^{(x)}}$ which maps a pure single-photon state with a Gaussian wavepacket given by the $(t,\omega)$ parameters to a coherent state $\ket{\alpha^{(x)}}$ with complex amplitude 
\begin{equation}
    \alpha^{(x)} = \frac{(t-x)}{2\sigma_t}+i\omega \sigma_t.
\end{equation}
The $\text{TF}^{(x)}_{\alpha}$ correspondence defines a vacuum-equivalent shot noise renormalized by $\sigma_t$ and $1/(2\sigma_t)$ for time and frequency, respectively. In the present context, the equivalent TF vacuum state may thus be associated to a Gaussian wavepacket, with $t=x$ and $\omega=0$. The equivalent coherent-state amplitude is therefore not an intrinsic property of the photon as it is defined relative to the detector's position $x$ defining the reference frame. In this light, changing the position reference frame $x_A\rightarrow x_B$ is equivalent to the change $\alpha^{(x_A)}\rightarrow \alpha^{(x_A)}-\delta_{BA}$, where $\delta_{BA}=(x_B-x_A)/(2\sigma_t)$. Consequently, since coherent states satisfy, up to a phase, $\hat D(\delta)\ket{\alpha} = \ket{\alpha + \delta}$, where $\hat D$ is the displacement operator in quadrature phase space, we have 
\begin{equation}
    \text{TF}^{(x_B)}_{\alpha} = \hat D(-\delta_{BA})\text{TF}^{(x_A)}_{\alpha}.
\end{equation}
In other words, space-time translations are represented by displacements in the effective quadrature phase space.
From here, the derivation of the CFT $\max \mathcal F_{\text{class}}$, which we derive step by step in Sup. Mat.~\textcolor{blue}{IV}, follows immediately from~\cite{braunstein2000criteria}, and we compute
\begin{equation}
\label{Equation measure-and-prepare CFT}
    \max \mathcal F_{\text{class}} = \frac{1}{2}.
\end{equation}
Consequently, the CFT for Gaussian single-photon TF states coincides with the result of~\cite{braunstein2000criteria} for coherent states in quadrature CV quantum teleportation, despite the fundamentally different physical encoding. The invariance of the CFT under detector translations is not to be confused with the teleportation fidelity $\mathcal F_{\text{Gaussian}}$ [Eq.~\eqref{Equation fidelity for a Gaussian wavepacket with an ideal quantum channel}], which corresponds to the transfer of a fixed physical Gaussian wavepacket between detector frames that have not been previously synchronized. In contrast, the benchmark presented here is an average over an ensemble of Gaussian wavepacket input states. \\
Furthermore, an interesting observation is that the CFT is independent of the wavepacket width $\sigma_t$. This invariance follows from the assumption of an ideal measurement, whose positive operator-valued measure (POVM) is defined with projectors onto the same Gaussian wavepackets as those comprising the input ensemble. More realistic detector models incorporating finite temporal resolution or frequency bandwidth~\cite{van2017time} are expected to curtail the maximum fidelity actually attainable.
\subsubsection{Best-Copy Teleportation Threshold}
\label{Section Best-copy teleportation threshold}
As previously mentioned, for quadrature CV quantum teleportation, the measure-and-prepare CFT is subsumed by the $\text{$1$-to-$M$}$ cloning threshold~\cite{grosshans2001quantum} in the limit of $M\rightarrow +\infty$ copies, while the best-copy threshold for $M=2$ is equal to $\max \mathcal F_{1\rightarrow 2} = 2/3$ for a Gaussian cloner. \\
For single-photon TF CVs, a similar threshold can be constructed building on the underlying Heisenberg algebra with the associated CCR $[\hat t_s,\hat \omega_s] = i$ in the single-photon subspace. The core argument of~\cite{grosshans2001quantum} for quadrature CVs can then be invoked in a similar fashion. Namely, consider a $1\rightarrow 2$ cloner with one input auxiliary mode $\text{in}$ and two output auxiliary modes $a$ and $b$. Assuming a unit gain, 
\begin{align}
    \hat t_a &= \hat t_{\text{in}}+\hat B_{\hat t_a}, \; \; \hat \omega_a = \hat \omega_{\text{in}}+\hat B_{\hat \omega_a} \\
    \hat t_b &= \hat t_{\text{in}}+\hat B_{\hat t_b}, \; \; \hat \omega_b = \hat \omega_{\text{in}}+\hat B_{\hat \omega_b},
\end{align}
where the $\hat B$s denote the noise added by the cloning channel. As detailed in Sup. Mat.~\textcolor{blue}{V}, because $[\hat t_a,\hat \omega_b] = 0$ for different auxiliary modes, assuming the output noises are not correlated with the input signal, the Robertson uncertainty relation~\cite{hall2013quantum} imposes a lower bound $\Delta \hat B_{\hat t_a}\Delta \hat B_{\hat \omega_b} \geq 1/2$ and $\Delta \hat B_{\hat t_b}\Delta \hat B_{\hat \omega_a} \geq 1/2$. \\
For a single-photon Gaussian wavepacket $\varphi_{t_0,\omega_0}(t)= \left(2\pi\sigma_t^2\right)^{-1/4}\;e^{-\frac{(t-t_0)^2}{4\sigma_t^2}}e^{i\omega_0(t-t_0)}$, this translates to the fidelity, derived in Sup. Mat.~\textcolor{blue}{V},
\begin{equation}
\label{Equation noisy copy fidelity}
    \mathcal F_{1\rightarrow 2} = \frac{1}{\sqrt{\left(1+\frac{\eta_t^2}{2\sigma_t^2}\right)\left(1+2\sigma_t^2\eta_\omega^2\right)}},
\end{equation}
where the $1\rightarrow 2$ copying procedure is described by a noisy Gaussian displacement of the input state. The extra TF noises $\eta_t$ and $\eta_\omega$ can, for instance, result from dispersion or the detectors' jitters. Assuming two identical copies, the Robertson lower bound implies $\eta_t \eta_\omega \geq 1/2$, hence the best-copy fidelity threshold
\begin{equation}
\label{Equation best-copy threshold}
    \max  \mathcal F_{1\rightarrow 2} = \frac{2}{3}
\end{equation}
obtained for $\eta_t^2 = \sigma_t^2$ and $\eta_\omega^2 = 1/(4\sigma_t^2)$. Having a fidelity higher than $2/3$ guarantees that no better copy of the target state is kept by either a cheating Alice or an eavesdropping Eve~\cite{grosshans2001quantum}. More specifically on the latter, an input-state-dependent calibration of the channel losses -- equivalently modeled as a TF noise -- is therefore needed in order to prevent an eavesdropping Eve from recovering a better copy than Bob's through access to the classical channel and through leaked information about the state. \\
The best-copy threshold Eq.~\eqref{Equation best-copy threshold} is equal to that of coherent states for quadrature CVs~\cite{grosshans2001quantum}. Unlike the CFT [Eq.~\eqref{Equation measure-and-prepare CFT}] derived in the previous section, the best-copy threshold does not hinge on a quadrature coherent state/single-photon Gaussian wavepacket correspondence but on the more fundamental Heisenberg algebra permitted by the $[\hat t_s,\hat \omega_s] = i\mathds{I}_{s}$ CCR in the single-photon subspace, which takes the same form as for quadrature CVs. Nevertheless, while for quadrature CVs the best-copy threshold for coherent-state inputs inherits the vacuum-noise fluctuations, which are independent of the input state, here, the input-dependent standard deviation $\sigma_t$ surfaces. This reflects what has been argued in the previous section, where we stated that the $\text{TF}_\alpha^{(x)}$ correspondence prompts the definition of a vacuum-equivalent shot noise renormalized by $\sigma_t$. As a consequence, for single-photon Gaussian wavepackets, any prior noise calibration is dependent on the standard deviation $\sigma_t$ characterizing a specific class of Gaussian states used for quantum teleportation. \\ \\
In Appendix~\ref{Appendix Protocol calculations in chronocyclic phase space}, when considering an arbitrary Gaussian quantum channel and an input Gaussian wavepacket, we obtain an input-output fidelity Eq.~\eqref{Appendix Equation input-output fidelity for Gaussian wavepacket and general Gaussian quantum channel} identical to $\mathcal F_{1\rightarrow 2}$ [Eq.~\eqref{Equation noisy copy fidelity}] but where the extra noises $\eta_t^2$ and $\eta_\omega^2$ are replaced by the two-photon quantum channel TF correlations $\langle\Delta \hat t_-^2\rangle$ and $\langle \Delta \hat \omega_+^2\rangle $
\begin{equation}
\label{Equation input-output fidelity for Gaussian wavepacket and general Gaussian quantum channel}
   \mathcal F_{\text{Gauss. channel}} = \frac{1}{\sqrt{ \left(1+\frac{\langle \Delta \hat t_-^2\rangle}{2\sigma_t^2}\right)\left(1+2\sigma_t^2\langle \Delta \hat \omega_+^2\rangle\right)}},
\end{equation}
In Eq.~\eqref{Equation input-output fidelity for Gaussian wavepacket and general Gaussian quantum channel}, we consider, for simplicity, that the detector-position mismatch has been corrected, as mentioned in Appendix~\ref{Appendix Protocol calculations in chronocyclic phase space}. The input-output fidelity [Eq.~\eqref{Equation input-output fidelity for Gaussian wavepacket and general Gaussian quantum channel}] is determined by the quantum-channel time- and frequency-noise variances relative to those of the input Gaussian wavepacket. For equal relative time- and frequency-noise contributions, namely $\frac{\langle \Delta \hat t_-^2\rangle}{2\sigma_t^2}=2\sigma_t^2\langle \Delta \hat\omega_+^2\rangle$, the classical fidelity threshold $\max\mathcal F_{\text{class}}=1/2$ demonstrated in Sec.~\ref{Section Measure-and-Prepare Strategy Threshold} is surpassed when $\langle \Delta \hat t_-^2\rangle<2\sigma_t^2$ and $\langle \Delta \hat\omega_+^2\rangle<\frac{1}{2\sigma_t^2}$. Therefore, the TF two-photon correlations required to surpass the classical fidelity threshold are set by the temporal width $\sigma_t$ of the input wavepacket. In contrast, for quadrature-CV coherent states, the fidelity and the corresponding condition for surpassing the classical threshold are expressed in terms of the quantum channel correlations $\langle \Delta \hat x_-^2\rangle$ and $\langle \Delta \hat p_+^2\rangle$ relative to the vacuum fluctuations, which provide a state-independent noise scale. No analogous state-independent vacuum-noise scale exists in the single-photon TF subspace, and the relevant reference scale is set by the input wavepacket through $\sigma_t$ and $1/(2\sigma_t)$ instead. \\
The characterization of the two-photon quantum channel correlations hinges on the reconstruction of the two-photon joint spectral amplitude, which has been extensively investigated~\cite{gianani2020measuring}, particularly for photon pairs generated by spontaneous parametric down-conversion (SPDC) sources. These approaches include stimulated-emission tomography~\cite{liscidini_stimulated_2013,eckstein2014high,Jizan:16,keller2022reconstructing}, which exploits the correspondence between SPDC and difference-frequency generation (DFG); intensity interferometry~\cite{thekkadath2022measuring}; and a phase-retrieval technique based on successive intensity measurements in Fourier-related domains combined with a reconstruction algorithm~\cite{PhysRevA.100.033834}.
\section{Discussion: Comparison with Quadrature-Based Continuous Variables}
CV quantum teleportation is typically associated with the position and momentum quadratures $\hat x$ and $\hat p$ of a bosonic mode, together with squeezed-vacuum states and homodyne detection~\cite{CVQTeleportationpirandola_quantum_2006,CVQTeleportationbraunstein1998teleportation}. The protocol we propose for TF CV quantum teleportation of single photons also operates in a CV framework, but differs substantially from conventional approaches in several key respects, which we discuss below. 
\label{Section Discussion}
\subsection{Physical System and the Role of Nonlinearities}
In our work, the CVs under consideration are the time of arrival and frequency of a \textit{single photon}, with the state restricted to the single-photon subspace. This encoding is therefore distinct from both conventional quadrature-based CV encodings, which rely on the photon-number statistics of a given mode, and DV encodings such as polarization, which exploit the discrete degrees of freedom of a single photon. \\ The quadrature-based CVs and the single-photon TF CVs are both underpinned by a Heisenberg algebra subsumed by their respective CCRs $[\hat x_s, \hat p_s] = i$ and $[\hat t_s,\hat \omega_s]=i$, the latter valid in the single-photon subspace~\cite{TFCVfabre2022time}. This allows the Braunstein-Kimble architecture~\cite{CVQTeleportationbraunstein1998teleportation} and associated phase-space methods~\cite{CVQTeleportationpirandola_quantum_2006} to be adapted to the TF degree of freedom, although the underlying physics differs considerably.  \\ 
The clearest difference is the role of nonlinearity. 
For quadrature-based CVs, the mode mixing is implemented using a balanced beamsplitter. The corresponding beamsplitter interaction is described by a quadratic Hamiltonian $\hat H_{\text{BS}} \sim \hat a^\dagger \hat b +\text{h.c.}$ and can be carried out using linear optics, while nonlinear optics is required only to prepare the squeezed-state quantum channel. In contrast, for TF CVs, the analogous mode-mixing operation is described~\cite{TFCVfabre2022time} by a quartic Hamiltonian $\hat H_{\text{TFBS}}\sim \iint dtdt'\;\hat a^\dagger\left(\frac{t+t'}{\sqrt{2}}\right)\hat b^\dagger\left(\frac{t-t'}{\sqrt{2}}\right)\hat a(t)\hat b(t') + \text{h.c.}$, so nonlinearity enters the Bell measurement itself, in addition to the two-photon time-frequency-entangled quantum channel. Although the present protocol concerns single-photon CVs, this requirement is consistent with standard DV single-photon quantum teleportation~\cite{nogothelinearcalsamiglia2001maximum,Generalhu2023progress}, where \textit{active} linear optics~\footnote{The authors in~\cite{ZaidiPhysRevLett.110.260501} refer to \textit{active} linear optics instead of nonlinear optics, but although squeezing operations correspond to interaction Hamiltonians that are quadratic in the bosonic operators and therefore implement linear Bogoliubov transformations~\cite{Generalpirandola2015advances,ContextCVserafini2023quantum}, they still require nonlinear interactions.} are required~\cite{KimPhysRevLett.86.1370,ZaidiPhysRevLett.110.260501} to enable complete Bell-state measurements and circumvent the quantum teleportation 50\% success-probability upper bound. Exceeding this upper bound with implementations based solely on \textit{passive} linear optics requires additional ancilla photons~\cite{bayerbach2023bell} or hyperentanglement~\cite{schuck2006complete}. \\ 
We further note that the single-photon frequency spectrum teleportation proposed by Humble in~\cite{humble_spectral_2010}, which thus manipulated a continuous degree of freedom, also requires nonlinearities. Specifically, the proposal in~\cite{humble_spectral_2010} relies on sum-frequency generation (SFG) between Alice's photon and the target photon to generate a new photon onto which a projective frequency measurement is then performed. In that regard, the SFG process considered in~\cite{humble_spectral_2010} does not constitute a genuine two-photon gate analogous to the balanced beamsplitter used for quadrature CVs, but rather maps the frequency information of the two photons onto a newly generated photon for subsequent measurement. In our approach, by contrast, the nonlinear TFBS realizes the two-photon mode-mixing interaction between the photons, playing the role of a genuine beamsplitter. Furthermore, platform proposals such as the frequency entangling gate (FrEnGATE)~\cite{FBSmeguebel2025generation} bypass the need for efficient nonlinear crystals by exploiting effective quadratic interactions mediated by the four-level structure of quantum dots embedded in waveguide structures, rather than relying on the intrinsic -- and typically weak -- nonlinear response of the optical medium.
\subsection{Reference Frame}
Another distinction concerns the treatment of reference frames. In our relational description of time, the time-of-arrival degree of freedom is defined conditionally on a detector position, and the light-cone constraint consequently makes this position part of the physical reference structure of the temporal degree of freedom. The resulting fidelity [Eq.~\eqref{Equation teleportation fidelity for ideal channels and measurements}] penalty from a detector-position mismatch is therefore not a defect of the quantum channel, but a structural consequence of the relational definition of the degree of freedom. In the standard quadrature-CV description, the phase reference with respect to which the quadratures are defined is implicitly taken as part of the background of the description. This assumption is precisely what Rudolph and Sanders~\cite{Referenceframerudolph2001requirement} questioned in their analysis of quadrature-CV quantum teleportation, which we discuss in Appendix~\ref{Appendix Section Superselection rules and reference externalization}. Formally, the demonstrated fidelity penalty [Eq.~\eqref{Equation teleportation fidelity for ideal channels and measurements}] and more specifically for an input Gaussian wavepacket [Eq.~\eqref{Equation fidelity for a Gaussian wavepacket with an ideal quantum channel}] is analogous to the fidelity loss produced by a \textit{displaced} quantum channel in quadrature CV teleportation (Sec.~IV.B. of~\cite{CVQTeleportationpirandola_quantum_2006}), but the origin of the displacement is fundamentally different: in our work, it is a structural consequence of how the time-of-arrival degree of freedom is defined.\\
This reference-frame dependence also echoes the distinction between \textit{speakable} and \textit{unspeakable} information addressed by Chiribella \textit{et al.}~\cite{Referenceframechiribella2012teleportation}. Speakable information (SI) can be specified independently of a reference frame, whereas unspeakable information (UI) is intrinsically defined relative to one and requires a shared reference to be faithfully interpreted. In our setting, the photon's temporal mode $\varphi(t)$ constitutes the SI, while its interpretation as a time-of-arrival degree of freedom depends on the detector-position reference, providing the corresponding UI. Chiribella \textit{et al.} showed that teleportation of both SI and UI requires prior synchronization of the relevant reference frame. Our result exhibits this same structural feature through the residual detector-position displacement. We stress, however, that their no-go theorem concerns reference frames associated with compact Lie groups, whereas the detector-position reference considered here is associated with the non-compact translation group on the real line. The SI/UI correspondence should therefore be understood as a structural analogy rather than as a direct application of their theorem. 
\subsection{Noise Scale}
A final comparison between quadrature CV quantum teleportation and single-photon TF CV quantum teleportation concerns the corresponding fidelity thresholds $\max\mathcal F_{\mathrm{class}}=1/2$ and $\max\mathcal F_{1\rightarrow2}=2/3$. Indeed, they coincide despite the very different physical systems involved, a single-photon wavepacket for TF CV quantum teleportation and a coherent state for quadrature CV quantum teleportation. This agreement follows from the correspondence between coherent states and single-photon Gaussian wavepackets, together with the TF Heisenberg algebra in the single-photon subspace~\cite{TFCVfabre2022time}. In this view, the best-copy fidelity threshold for an optimal \textit{non-Gaussian} cloner of approximately 0.6826 derived in~\cite{cerf2005non} for quadrature CVs can equivalently be computed, as the calculations therein are formulated in phase space which carries over to the single-photon TF CVs with the associated chronocyclic phase space~\cite{BrechtPhysRevA.87.053810,fabre2020generation,TFCVfabre2022time}. However, what still differs between the two CV teleportation schemes is the role of the vacuum reference. Unlike in quadrature-based CVs, there is no input-independent vacuum-noise scale for single-photon TF CVs. Instead, the relevant vacuum-equivalent scale is set by the input wavepacket itself, making the noise tolerance, and hence the certification of quantum teleportation, dependent on the chosen input alphabet.

\section{Conclusion}
\label{Section Conclusion}
We have introduced a single-photon TF CV quantum teleportation protocol rooted in a relational treatment of the time-of-arrival degree of freedom, in which time is defined conditionally on a detector position. We demonstrate that the protocol teleports both the photon's temporal wavefunction and the reference with respect to which its time of arrival is defined. As a result, even for an ideal two-photon quantum channel and perfect classical communication, a detector-position mismatch produces a fidelity penalty set by the temporal width of the input wavepacket. This establishes a fidelity/temporal-bandwidth trade-off, where the narrower the photon in time, the more sensitive the teleportation fidelity is to the relative detector-position reference. Crucially, Alice and Bob need not be at the same location because, once the relative detector position is known, the corresponding temporal displacement can be corrected. Moreover, the protocol itself can be used to calibrate this reference. Indeed, a known input state can be teleported first, allowing the correction to be optimized and the relative detector position to be remotely synchronized. TF cat and GKP states provide particularly structured probes through their fidelity revivals, which encode internal temporal scales of the states. \\
We have further established two quantumness benchmarks for single-photon Gaussian wavepackets: a classical-limit fidelity threshold $\max \mathcal F_{\mathrm{class}}=1/2$ for measure-and-prepare strategies and a best-copy threshold $\max \mathcal F_{1\rightarrow2}=2/3$ for a Gaussian cloner. Their coincidence with the corresponding benchmarks for quadrature-CV coherent states follows from the correspondence between single-photon Gaussian wavepackets and coherent states together with the underlying TF Heisenberg algebra~\cite{braunstein2000criteria,grosshans2001quantum,TFCVfabre2022time}. At the same time, the associated noise scale is not input-independent: in the single-photon TF subspace, the vacuum-equivalent reference is set by the input state. \\ 
Our analysis assumes ideal projective TF measurements, lossless channels and a perfect nonlinear TFBS. As discussed, timing jitter, finite spectral resolution and delay fluctuations act as additional TF noise in practice. Quantifying these effects further is a natural next step. Sub-nanometer spectral and picosecond temporal resolutions have been demonstrated simultaneously in the grating-based single-photon spectrometer presented in~\cite{jirsa2025fast}. An efficient nonlinear TFBS, in contrast, has yet to be demonstrated experimentally. One should note, however, that photons scattered off a quantum dot in a waveguide have been shown to display temporal quantum correlations~\cite{TBSle2022dynamical}, although through a linear light-matter coupling and with weak coherent states rather than single photons. \\
Finally, our construction of the photon's time-of-arrival degree of freedom differs from clock-based treatments~\cite{Quantumclockmaccone2020quantum,roncallo2023does}, in which time is carried by a clock, to whose energy it is conjugate, and the detector position enters as a projector on the system. In our work, the time of arrival is the quantum degree of freedom of the photon itself -- conjugate to its frequency -- and the detector position plays the role of the reference. This relational definition connects the single-photon time-frequency framework to quantum reference frames and superselection rules, with the detector externalized as a classical label, as is done for the reference mode in the particle-number case~\cite{Referenceframebartlett2007reference,Superselectionrulesdescamps2024superselection}. \\ \\
In summary, our work adds a quantum teleportation protocol to the growing toolbox of single-photon TF CV quantum information. The protocol provides a means to transfer single-photon TF states between distant nodes, the fidelity benchmarks needed to certify the transfer and the treatment of the detector reference that any networked implementation must include. Our proposal therefore opens the way towards applications built on quantum teleportation~\cite{Generalpirandola2015advances,Generalhu2023progress}, from quantum computing to quantum networks. \\
A natural extension of the present work is the teleportation of photonic states in which both the spectro-temporal degrees of freedom and the particle-number statistics are transferred. Beyond teleportation itself, the possibility of remotely establishing a common detector-position reference using the teleportation protocol suggests a complementary route to remote synchronization~\cite{giovannetti2002positioning,ho2009clock,spiess2023clock}. Such a capability is relevant to distributed quantum communication architectures, including twin-field quantum key distribution, where independent remote sources must maintain stable temporal and phase references for interference at a central station~\cite{lucamarini2018overcoming,li2026free}. \\

\section*{ACKNOWLEDGMENT}
The present work received funding from the French State, managed by Agence Nationale de la Recherche under the France 2030 program, and from the European Union through NextGenerationEU, under the reference ANR-22-CMAS-0001 France 2030. N. Fabre and M. Meguebel acknowledge fruitful discussions with Eloi Descamps, Arne Keller and Pérola Milman.  

\section*{Disclosures}
Artificial intelligent agent \textit{Claude Sonnet 5.5} was used exclusively for checking spelling, typos and grammar. All scientific content was verified by the authors, who take full responsability for the manuscript. \\ \\
The authors declare no conflicts of interest.

\section*{Data availability}

Data underlying the results presented in this paper are not publicly available at this time but may be obtained from the authors upon reasonable request.

\onecolumngrid

\appendix

\section{Quantum Teleportation General Output State and Input-Output Fidelity: Calculations in Chronocyclic Phase Space}
\label{Appendix Protocol calculations in chronocyclic phase space}
In this appendix, we derive the general output state [Eq.~\eqref{Appendix Equation general average output state}] at Bob's side. Notably, we provide a simple form for the input-output fidelity calculated for a pure Gaussian state and a general Gaussian two-photon quantum channel. Our calculations are performed in the chronocyclic phase space~\cite{BrechtPhysRevA.87.053810,fabre2020generation,TFCVfabre2022time}, analogously to the quadrature phase space. \\ \\
The input state $\hat{\rho}_{\text{in}}$ to be teleported can be expressed as 
\begin{equation}
    \hat \rho_{\text{in}} = \int dtdt'\; \rho_{\text{in}}(t,t')\ket{t}\bra{t'}.
\end{equation}
Since we will be focusing on Gaussian elements in the end, the calculations are conveniently performed in phase space with the Wigner function~\cite{CVfieldquadraturebraunstein2005quantum,ContextCVserafini2023quantum}, and in particular the chronocyclic Wigner function~\cite{paye1992chronocyclic}. More specifically, in the single-photon subspace~\cite{BrechtPhysRevA.87.053810,fabre2020generation,TFCVfabre2022time}, we define the chronocyclic Wigner  function $W_{\text{in}}(t,\omega)$ as a symplectic Fourier transform
\begin{equation}
\label{Appendix Equation input Wigner function with characterstic function}
    W_{\text{in}}(t,\omega) = \frac{1}{(2\pi)^2}\int_{\mathds R^2} dt'd\omega'e^{i(\omega't-\omega t')}\chi_{\text{in}}(t',\omega'),
\end{equation}
with $\chi_{\text{in}}(t',\omega')$ the symmetric characteristic function defined as \begin{equation}
    \chi_{\text{in}}(t',\omega') = \Tr[\hat \rho_{\text{in}}\hat{\mathcal D}_s^\dagger(t',\omega')]
\end{equation}
where $\hat{\mathcal D}_s(t',\omega') = e^{-i\omega't'/2}\hat{\mathcal D}_\omega(\omega')\hat{\mathcal D}_t(t')$ the symmetric displacement operator in the single-photon TF phase space~\cite{fabre2020generation}. We recall $\hat{\mathcal D}_{t}(\tau)\ket{t} = \ket{t+\tau}$ and $\hat{\mathcal D}_\omega(\nu)\ket{t} = e^{i\nu t}\ket{t}$ for time $\tau$ and frequency shifts $\nu$. The chronocyclic Wigner function Eq.~\eqref{Appendix Equation input Wigner function with characterstic function} can equivalently be expressed as
\begin{equation}
    W_{\text{in}}(t,\omega) = \frac{1}{2\pi}\int_{\mathds R}dt'e^{-i\omega t'}\langle t+\frac{t'}{2} | \hat \rho_{\text{in}} | t-\frac{t'}{2}\rangle.
\end{equation}
The chronocyclic phase space is built on the $[\hat t_s, \hat \omega_s] = i$ TF CCR in the single-photon subspace~\cite{TFCVfabre2022time}, where the time and frequency operators are defined, respectively, as $\hat t_s = \int dt\; t\ket{t}\bra{t}$ and $\hat \omega_s = \int d\omega\; \omega \ket{\omega}\bra{\omega}$. This CCR defines a Heisenberg algebra akin to that of quadratures, hence the immediate correspondence between the usual quadrature CV Wigner function and the present single-photon chronocyclic Wigner function. \\
The detector-position reference frame discussed in the main text is contained within the density matrix $\hat \rho_{\text{in}}$ and will reappear later when taking specific input states and quantum channels. \\
For photons occupying different auxiliary modes, the chronocyclic Wigner function extends immediately~\cite{fabre2020generation,TFCVfabre2022time}, and in particular for the two-photon quantum channel $\hat \rho_{AB}$ shared by Alice and Bob 
\begin{equation}
    W_{AB}(t_A,\omega_A,t_B,\omega_B) = \frac{1}{(2\pi)^2}\int_{\mathds R^2}d\tau d\tau' e^{-i\omega_A \tau}e^{-i\omega_B\tau'}\left\langle t_A+\frac{\tau}{2},t_B+\frac{\tau'}{2} \bigg|\hat \rho_{AB} \bigg| t_A-\frac{\tau}{2},t_B-\frac{\tau'}{2}\right\rangle.
\end{equation}
From here, most of the calculation follows Pirandola and Mancini's review~\cite{CVQTeleportationpirandola_quantum_2006} for quadrature CV quantum teleportation, see Sec.~IV.B.  Let us now go through the protocol steps (S1) to (S4) presented in the main text.  \\
The initial state is given by the tensor product $\hat \rho_{\text{in}}\otimes \hat \rho_{AB}$, in such a way that the initial chronocyclic Wigner function reads as a product $W(t,\omega,t_A,\omega_A,t_B,\omega_B) = W_{\text{in}}(t,\omega)W_{AB}(t_A,\omega_A,t_B,\omega_B)$. (S1) The mode mixing performed by the nonlinear TFBS leads to 
\begin{equation}
    W(t_+,\omega_+,t_-,\omega_-,t_B,\omega_B) = W_{\text{in}}\left(\frac{t_+-t_-}{\sqrt{2}},\frac{\omega_+-\omega_-}{\sqrt{2}}\right)W_{AB}\left(\frac{t_++t_-}{\sqrt{2}},\frac{\omega_++\omega_-}{\sqrt{2}},t_B,\omega_B\right).
\end{equation}
Indeed, if the phase-space coordinates $\boldsymbol r$ are modified according to $\boldsymbol r' = M\boldsymbol r$ where $M$ is symplectic, then the Wigner function transforms as $W(\boldsymbol r) \rightarrow W(M^{-1}\boldsymbol r')$, see Sec.~II. A. of~\cite{CVQTeleportationpirandola_quantum_2006}. \\
Next, (S2) Alice measures the TF quadratures $\hat t_-$ and $\hat \omega_+$ jointly, since they commute, with the outcomes $t_-$ and $\omega_+$. Projective measurements of the $\hat t_-$ and $\hat \omega_+$ quadratures in the chronocyclic Wigner function representation amount to fixing the measurement outcome and marginalizing over its associated conjugate variable. Therefore, Bob's conditional chronocyclic Wigner function reads 
\begin{equation}
    W_B(t_B,\omega_B |t_-,\omega_+) = \frac{1}{p(t_-,\omega_+)}\int dt_+d\omega_-\; W_{\text{in}}\left(\frac{t_+-t_-}{\sqrt{2}},\frac{\omega_+-\omega_-}{\sqrt{2}}\right)W_{AB}\left(\frac{t_++t_-}{\sqrt{2}},\frac{\omega_++\omega_-}{\sqrt{2}},t_B,\omega_B\right),
\end{equation}
where $p(t_-,\omega_+)$ is the probability of measuring $t_-$ and $\omega_+$. Consequently, defining $\omega = (\omega_+-\omega_-)/\sqrt{2}$ and $t=(t_+-t_-)/\sqrt{2}$
\begin{equation}
    W_B(t_B,\omega_B |t_-,\omega_+) = \frac{2}{p(t_-,\omega_+)}\int dt d\omega\; W_{\text{in}}\left(t,\omega\right)W_{AB}\left(\sqrt{2}t_-+t,\sqrt{2}\omega_+-\omega,t_B,\omega_B\right).
\end{equation}
(S3) Alice then sends her measurement outcome to Bob. (S4) Bob then performs the conditional displacements $t_B \rightarrow t_B-\sqrt{2}t_-$ and $\omega_B \rightarrow \omega_B+\sqrt{2}\omega_+$ to correct for the shifts induced by the Bell measurement stage. In terms of the chronocyclic Wigner function $W_B(t_B,\omega_B |t_-,\omega_+) \rightarrow W_B(t_B+\sqrt{2}t_-,\omega_B-\sqrt{2}\omega_+ | t_-, \omega_+)$. The average chronocyclic Wigner function at Bob's side reads 
\begin{equation}
    W_B^{\text{av.}}(t_B,\omega_B) = \int dt_-d\omega_+p(t_-,\omega_+)W_B(t_B+\sqrt{2}t_-,\omega_B-\sqrt{2}\omega_+ | t_-, \omega_+),
\end{equation}
that is, 
\begin{equation}
\label{Appendix Equation general average output state}
    W_B^{\text{av.}}(t_B,\omega_B) = \int dtd\omega\; W_{\text{in}}(t,\omega)K(t,\omega,t_B,\omega_B),
\end{equation}
where, rescaling $t_-$ and $\omega_+$, we have $K(t,\omega,t_B,\omega_B)$ the distortion kernel
\begin{equation}
K(t,\omega,t_B,\omega_B) = \int dt_-d\omega_+\; W_{AB}\left(t_-+t,\omega_+-\omega,t_B+t_-, \omega_B-\omega_+\right),
\end{equation}
which depends on the quantum channel. \\ 
For a pure input state, the input-output fidelity reduces to 
\begin{equation}
\label{Appendix Equation general input-output fidelity for a pure state}
    \mathcal F = 2\pi\int dt d\omega\; W_{\text{in}}(t,\omega)W_B^{\text{av.}}(t,\omega).
\end{equation}
Let us now consider Gaussian input states and Gaussian quantum channels, characterized only~\cite{weedbrook2012gaussian,brask2021gaussian} by their first and second moments, that is to say the displacements $d_{\text{in}}$ and $d_{AB}$, and the correlation matrices $V_{\text{in}}$ and $V_{AB}$, respectively. In a similar fashion to~\cite{CVQTeleportationpirandola_quantum_2006} Sec.~IV.B, we obtain 
\begin{equation}
    \mathcal F = \frac{1}{\sqrt{\det \Gamma}}\exp\left(-\frac{1}{2}h^T\Gamma^{-1}h\right),
\end{equation}
where 
\begin{equation}
    \Gamma = 2 V_{\text{in}} + ZAZ + B-ZC -C^TZ^T,
\end{equation}
where $Z = \text{diag}(1,-1)$ and where we write $V_{AB}$ in the block form $V_{AB} = \begin{bmatrix}A & C \\ C^T & B \end{bmatrix}$, and with $h =(\langle \hat t_B-\hat t_A\rangle, \langle \hat \omega_A+\hat \omega_B \rangle)^T$. In turn, any shift $h$ inherited from the two-photon quantum channel inevitably leads to a Gaussian decay of the fidelity. Bob could in principle suppress this shift by applying an additional conditional displacement. Nonetheless, the shift $h$ contains the detector-position mismatch which cannot be accounted for without prior synchronization, as extensively argued in the main text. For the sake of argument, however, let us now consider the shift $h$ to be zero, resulting in a fidelity $\mathcal F = 1/\sqrt{\det \Gamma}$. \\
For a single-photon Gaussian wavepacket (without chirp), the correlation matrix reads $V_{\text{in}} = \text{diag}\left(\sigma_t^2,1/(4\sigma_t^2)\right)$. In turn, 
\begin{equation}
\label{Appendix Equation input-output fidelity for Gaussian wavepacket and general Gaussian quantum channel}
   \mathcal F = \frac{1}{\sqrt{ \left(1+\frac{\langle \Delta \hat t_-^2\rangle}{2\sigma_t^2}\right)\left(1+2\sigma_t^2\langle \Delta \hat \omega_+^2\rangle\right)}},
\end{equation}
with $\hat t_- = \hat t_A-\hat t_B$ and $\hat \omega_+ = \hat \omega_A+\hat \omega_B$. The input-output fidelity Eq.~\eqref{Appendix Equation input-output fidelity for Gaussian wavepacket and general Gaussian quantum channel} takes the same form as the fidelity $\mathcal F_{1\rightarrow 2}$ [Eq.~\eqref{Equation noisy copy fidelity}] obtained in Sec.~\ref{Section Best-copy teleportation threshold} from a noisy copy of the input state. A quantum teleportation protocol with an imperfect two-photon quantum channel, \textit{i.e.}, with $\langle \Delta \hat t_-^2\rangle\neq0$ and $\langle \Delta \hat\omega_+^2\rangle\neq0$, can thus be interpreted as a $1\rightarrow2$ copying procedure in which the additional TF noises are determined by the two-photon quantum channel correlations, $\langle \Delta \hat t_-^2\rangle$ and $\langle \Delta \hat\omega_+^2\rangle$. That said, unlike the copy-induced TF noises, $\langle \Delta \hat t_-^2\rangle$ and $\langle \Delta \hat \omega_+^2\rangle$ are not lower bounded, and the fidelity Eq.~\eqref{Appendix Equation input-output fidelity for Gaussian wavepacket and general Gaussian quantum channel} can in principle reach $\mathcal F = 1$ for an ideal EPR state with $\langle \Delta \hat t_-^2\rangle = 0$ and $\langle \Delta \hat \omega_+^2\rangle=0$. \\ \\
As previously argued, the input-output fidelity in Eq.~\eqref{Appendix Equation input-output fidelity for Gaussian wavepacket and general Gaussian quantum channel} highlights a key difference between quadrature-based CV quantum teleportation and single-photon TF CV quantum teleportation. For quadrature-CV coherent states, the fidelity and the corresponding condition for surpassing the classical threshold are determined by the quantum channel correlations $\langle \Delta \hat x_-^2\rangle$ and $\langle \Delta \hat p_+^2\rangle$ relative to the vacuum fluctuations, which provide a state-independent reference noise scale. On the other hand, for single-photon TF CV quantum teleportation the corresponding time- and frequency-noise variances are referenced to the input wavepacket, through the scales $\sigma_t$ and $1/(2\sigma_t)$. 
\section{Single-Photon Time-Frequency Continuous-Variable Quantum Teleportation in the Context of Superselection Rules and Reference Externalization}
\label{Appendix Section Superselection rules and reference externalization}
In Sec.~\ref{Section Quantum teleportation protocol} of the main text, we discuss the requirement to explicitly teleport the detector reference frame. In this Appendix, we exhibit how this echoes Rudolph and Sanders'~\cite{Referenceframerudolph2001requirement} analysis of the intrinsic optical coherence requirement in quadrature-based continuous-variable (CV) quantum teleportation. 
\subsection{Particle-Number Superselection Rule} 
Rudolph and Sanders' rationale is that -- contrary to the standard practice in quantum optics -- one should not assume that an optical mode $S$ can be legitimately described by a coherent superposition of Fock states belonging to different total photon-number sectors. That is, the usual expression $\ket{\psi}_S = \sum_{n \in \mathds N}c_n\ket{n}_S$ would be disallowed under a particle-number superselection rule (SSR), because it involves coherence between different number sectors. In their view, an SSR-compliant description must include an appropriate reference-frame mode $R$. Concretely, rather than the state $\ket{\psi}_S$, one should instead use an SSR-compliant state $\ket{\psi}_{\text{SSR}}$~\cite{Referenceframebartlett2007reference,Superselectionrulesdescamps2024superselection}
\begin{equation}
\label{Equation SSR-compliant coherent state}
\ket{\psi}_S \leftarrow \ket{\psi}_{\text{SSR}}=\sum_{n = 0}^{N}c_n\ket{n}_S\otimes \ket{N-n}_R ,
\end{equation}
where $N$ is the fixed total photon number shared by modes $S$ and $R$. In quadrature-based CV quantum teleportation, the reference mode $R$ that one would need -- in order to give physical meaning to an SSR-compliant coherent state -- would usually correspond to the physical laser medium, which is inaccessible in practice. Consequently, one cannot meaningfully assert that the mode sent through the teleportation channel is in a pure coherent state. Rather, lacking access to the reference, the state must be described as a statistical mixture over unknown global phases~\cite{Referenceframerudolph2001requirement}. Under this SSR description, Rudolph and Sanders argued that the standard CV teleportation protocol loses its claim to teleport a pure coherent state~\cite{Referenceframerudolph2001requirement}. Rudolph and Sanders' position is controversial~\cite{Referenceframewiseman2003comment,CVfieldquadraturebraunstein2005quantum,bartlett2006dialogue}, but it provides an interesting relation to our work, where the single-photon time-frequency (TF) CV quantum teleportation is rooted in the explicit introduction of a detector reference frame giving meaning to the time-of-arrival degree of freedom.  \\
\subsection{Causal Superselection-like Rule}
In our work, a causal SSR-like rule emerges through the d'Alembertian operator $\Box \ket{\psi} = 0$ imposed by the light-cone constraint presented in Sec.~\ref{Section Light-cone constraint}. In terms of energy-momentum event, this constraint becomes a quadratic cross-system relation between the photon energy and the detector momentum: $E^2 = c^2 p_x^2$, akin to the particle-number SSR $n_S+n_R = N$, as expressed in Eq.~\eqref{Equation SSR-compliant coherent state}. The essential distinction between the particle-number SSR and the causal SSR-like rule in the context of CV teleportation is that quadrature-based protocols lack practical access to the reference frame, associated with the laser medium for a coherent state. By contrast, for TF CV teleportation as presented in our work, the detector's position is available to the experimenter in practice. Therefore, externalizing~\cite{Referenceframebartlett2007reference,Superselectionrulesdescamps2024superselection} the reference is immediate and operationally sound, as it amounts to fixing the detector's position $x$ to a classical label $x_D$, that is $\ket{\psi(x)}_S\ket{x}_D\rightarrow \ket{\psi(x_D)}_S$ in Eq.~\eqref{Equation space-time foliation with detector} of the main text
\begin{equation}
     \ket{\psi}_{\text{det}} = \int_{\mathds R}dx\;\ket{\psi(x)}_S\otimes\ket{x}_D.
\end{equation}
Therefore, with the light-cone constraint for right-propagating photons discussed in Sec.~\ref{Section Light-cone constraint} 
\begin{equation}
     \ket{\psi}_{\text{det}} = \int_{\mathds R}dt\; \varphi(t-x_D)\ket{t}_S.
\end{equation}
TF CV teleportation is therefore not merely an analogue of quadrature CV teleportation with a different superselection structure. It is a framework in which the SSR constraint becomes explicit and experimentally tractable. Furthermore, resorting to the detector-conditioned Page-Wootters formulation is not solely of foundational interest regarding the nature of time in quantum mechanics. The construction also directly addresses a concrete ambiguity in CV teleportation protocols, namely the necessity and explicit treatment of reference frames. 

\bibliography{bibliocvteleportation}

\end{document}


\author{Mohamed Meguebel}
\email{mohamed.meguebel@telecom-paris.fr}
\affiliation{Telecom Paris, Institut Polytechnique de Paris, 19 Place Marguerite Perey, 91120 Palaiseau, France}
\author{Nadia Belabas}
\affiliation{Centre for Nanosciences and Nanotechnology, CNRS, Université Paris-Saclay, 10 Bd Thomas Gobert, 91120 Palaiseau, France}
\author{Nicolas Fabre}
\email{nicolas.fabre@telecom-paris.fr}
\affiliation{Telecom Paris, Institut Polytechnique de Paris, 19 Place Marguerite Perey, 91120 Palaiseau, France}

\date{\today}
\begin{abstract}
The present document is a supplementary material providing calculation details for the article \textbf{Time-Frequency Continuous-Variable Quantum Teleportation with Single Photons}. In Sec.~\ref{Appendix Light-cone coordinates and normalization} we provide a straightforward normalization prescription to a geometric event-based (GEB) state associated with a right-propagating photon. Next, in Sec.~\ref{Appendix Time-frequency cat state fidelity revivals} and Sec.~\ref{Appendix Time-frequency GKP state fidelity revivals}, we compute the normalization constants and input-output fidelities for pure time-frequency cat and Gottesman–Kitaev–Preskill (GKP) states. These input-output fidelities are given for a quantum channel with EPR correlations $\phi(t_A,t_B) = \delta(t_A-t_B)$, as described in Sec.~\textcolor{blue}{III.C.1} of the main text. Finally, in Sec.~\ref{Appendix Measure-and-prepare strategy fidelity threshold derivation} and Sec.~\ref{Appendix Best-copy threshold derivation}, we derive, respectively, the measure-and-prepare and best-copy fidelity thresholds in our time-frequency continuous-variable quantum teleportation protocol for a single-photon Gaussian wavepacket.
\end{abstract}
\pacs{}
\vskip2pc 
\title{Time-Frequency Continuous-Variable Quantum Teleportation with Single Photons}
\maketitle
\onecolumngrid
\section{Light-cone coordinates and normalization}
\label{Appendix Light-cone coordinates and normalization}
This section is dedicated to the normalization of the light-cone-constrained GEB state
\begin{equation}
\label{Appendix Equation GEB state under light-cone constraint}
    \ket{\psi} = \int_{\mathbb R^2}dtdx\; \varphi\left(t-x\right)\ket{t,x},
\end{equation}
from Sec~\textcolor{blue}{II.B} of the main text. The state Eq.~\eqref{Appendix Equation GEB state under light-cone constraint} is ill-defined. Indeed, consider the light-cone coordinates 
\begin{equation}
    u = \frac{t-x}{\sqrt{2}}, \; \; v = \frac{t+x}{\sqrt{2}}
\end{equation}
with in turn
\begin{equation}
     \braket{\psi} = \int_{\mathbb R}du\; |\varphi(u)|^2\int_{\mathbb R}dv,
\end{equation}
which correspond to an unphysical state. This normalization issue originates from embedding a two-dimensional space-time manifold into a one-dimensional degree of freedom $u=t-x$. The amplitude $\varphi(u)$ should therefore not be interpreted as a square-integrable wavefunction over the full $(t,x)$ space-time, but rather as a wavefunction defined on the coordinate $u$, that is on the family of light cones labelled by $u$. We thus prescribe the normalization condition 
\begin{equation}
\int_{\mathbb R}du\; |\varphi(u)|^2 = 1.
\end{equation}
The divergence arising from the conjugate coordinate $v$ can be regularized by restricting $v$ to a finite interval of length $L$. The regulated norm then reads 
\begin{equation}
\braket{\psi}_{\text{reg}} = \int_{\mathbb R}du\;|\varphi(u)|^2\int_{-L/2}^{L/2}dv = L\int_{\mathbb R}du\;|\varphi(u)|^2.
\end{equation}
The physical norm is subsequently obtained by dividing out this divergent length
\begin{equation}
\label{Equation state normalization prescription}
\braket{\psi}_{\text{phys}} = \lim_{L\rightarrow +\infty}\frac{1}{L}\braket{\psi_\text{reg}} = \int_{\mathbb R}du\; |\varphi(u)|^2.
\end{equation}
This procedure identifies the physically relevant Hilbert space with $L^2(\mathbb R,du)$, corresponding to states defined on the coordinate $u$ rather than on the full space-time manifold, as enforced by the light-cone constraint for a right-propagating photon.
\section{Time-frequency cat state fidelity revivals}
\label{Appendix Time-frequency cat state fidelity revivals}
In this section, we derive the normalization constant $\mathcal N_{\text{cat}}$ along with the fidelity $\mathcal F_{\text{cat}}$ for pure time-frequency cat states, as presented in the main text in Sec.~\textcolor{blue}{III.C.3}. \\
Define the time-frequency cat state $\ket{\psi_{\text{TF,cat}}} = \mathcal N_{\text{cat}}\left(\ket{t_0,\omega_0}+e^{i\theta}\ket{-t_0,-\omega_0}\right)$, where $\ket{t_0,\omega_0}$ corresponds to a single-photon pure state with a Gaussian wavepacket \begin{equation}
    \ket{t_0,\omega_0} = \frac{1}{(2\pi s_t^2)^{1/4}}\int dt\; e^{-\frac{(t-t_0)^2}{4s_t^2}}e^{i\omega_0(t-t_0)}\ket{t}
\end{equation}
of positive parameters $t_0$ and $\omega_0$. Note that the central frequency of a single-photon wavepacket should be positive. Here, one of the wavepackets has a central frequency of $-\omega_0$ instead. This is justified by defining the central frequencies with respect to an arbitrary reference frequency $\omega_{\text{ref}}$, such that they are understood as frequency detunings relative to $\omega_{\text{ref}}$ rather than as absolute frequencies. \\
$\ket{\psi_{\text{TF,cat}}}$ is normalized to 1, as follows 
\begin{equation}
    |\mathcal N_{\text{cat}}|^4\left(2\Re e^{i\theta}\langle -t_0,-\omega_0 | t_0,\omega_0\rangle+2\frac{1}{(2\pi s_t^2)^{1/2}}\int dt\; e^{-\frac{t^2}{2s_t^2}}\right)^2 = 1.
\end{equation}
Using $\int dt\;e^{-\frac{1}{2}at^2+bt}=\left(2\pi/a\right)^{1/2}e^{\frac{b^2}{2a}}$ for $a,b\in \mathbb C$ and $\Re a > 0$, 
\begin{equation}
    \mathcal N_{\text{cat}} = \frac{1}{\sqrt{2\left(1+\cos\theta e^{-\frac{t_0^2}{2s_t^2}-2s_t^2\omega_0^2}\right)}},
\end{equation}
where we chose $\mathcal N_{\text{cat}}$ to be real positive without loss of generality. The normalization constant $\mathcal N_{\text{cat}}$ takes the same form as for quadrature cat states constructed from coherent states, and therefore warrants the definition 
\begin{equation}
    |\alpha|^2 = \frac{t_0^2}{4s_t^2}+s_t^2\omega_0^2.
\end{equation}
The input state to be teleported by Alice to Bob reads 
\begin{equation}
    \ket{\psi}_{\text{in}} = \int_{\mathbb R}dt\;\varphi_{\text{cat}}(t)\ket{t+x_{\text{in}}}_{\text{in}},
\end{equation}
with $ \varphi_{\text{cat}}(t) = \langle t | \psi_{\text{TF,cat}} \rangle$. As demonstrated in the main text, for an ideal quantum channel the input-output state fidelity is given as $ \mathcal F_{\text{cat}} = \left|\int_{\mathbb R}dt\; \varphi_{\text{cat}}(t)\varphi^*_{\text{cat}}\left(t+\Delta x_{BA}\right)\right|^2$ with $\Delta x_{BA}$ the detector position mismatch. In turn
\begin{equation}
\begin{split}
    \mathcal F_{\text{cat}} &= |\mathcal N_{\text{cat}}|^4\bigg|2\cos(\omega_0\Delta x_{BA}) e^{-\frac{\Delta x_{BA}^2}{8s_t^2}}\\
    &+\frac{1}{(2\pi s_t^2)^{1/2}}\left(e^{-i(\theta-\omega_0\Delta x_{BA})}\int dt\; e^{2i\omega_0t}e^{-\frac{(t-t_0)^2}{4s_t^2}}e^{-\frac{(t+t_0+\Delta x_{BA})^2}{4s_t^2}}+e^{i(\theta-\omega_0\Delta x_{BA})}\int dt\; e^{-2i\omega_0t}e^{-\frac{(t+t_0)^2}{4s_t^2}}e^{-\frac{(t-t_0+\Delta x_{BA})^2}{4s_t^2}} \right)\bigg|^2.
\end{split}
\end{equation}
Expressing $(t+a)^2+(t+b)^2 = c(t+\tau)^2+d$, where $c=2$, $\tau = (a+b)/2$ and $d=(a-b)^2/2$ and using the previous Gaussian integral identity, we obtain 
\begin{equation}
    \mathcal F_{\text{cat}} = |\mathcal N_{\text{cat}}|^4\left|2\cos(\omega_0\Delta x_{BA}) e^{-\frac{\Delta x_{BA}^2}{8s_t^2}}+e^{-i\theta} e^{-\frac{(\Delta x_{BA}+2t_0)^2}{8s_t^2}}e^{-2\omega_0^2s_t^2}+e^{i\theta}e^{-\frac{(\Delta x_{BA}-2t_0)^2}{8s_t^2}}e^{-2\omega_0^2s_t^2}\right|^2,
\end{equation}
that is 
\begin{equation}
    \mathcal F_{\text{cat}} = \frac{\left|2\cos(\omega_0\Delta x_{BA}) e^{-\frac{\Delta x_{BA}^2}{8s_t^2}}+e^{-i\theta} e^{-\frac{(\Delta x_{BA}+2t_0)^2}{8s_t^2}}e^{-2\omega_0^2s_t^2}+e^{i\theta}e^{-\frac{(\Delta x_{BA}-2t_0)^2}{8s_t^2}}e^{-2\omega_0^2s_t^2}\right|^2}{4\left(1+\cos\theta e^{-2|\alpha|^2}\right)^2}.
\end{equation}
For well separated lobes, $t_0\gg s_t$ and
\begin{equation}
    \mathcal F_{\text{cat}} \approx \cos^2(\omega_0\Delta x_{BA})e^{-\frac{\Delta x_{BA}^2}{4s_t^2}}+\frac{1}{4}e^{-\frac{(\Delta x_{BA}+2t_0)^2}{4s_t^2}}e^{-4\omega_0^2s_t^2}+\frac{1}{4}e^{-\frac{(\Delta x_{BA}-2t_0)^2}{4s_t^2}}e^{-4\omega_0^2s_t^2},
\end{equation}
that is with a central peak at zero detector mismatch $\Delta x_{BA} = 0$ and two partial revivals at $\Delta x_{BA} = \pm 2t_0$ where the fidelity reaches $\mathcal F_{\text{cat}} \approx \frac{1}{4}e^{-4\omega_0^2s_t^2}$.
\section{Time-frequency Gottesman–Kitaev–Preskill state fidelity revivals}
\label{Appendix Time-frequency GKP state fidelity revivals}
In this section, we derive the normalization constants $\mathcal N_{\text{even}}$ and $\mathcal N_{\text{odd}}$ along with the corresponding fidelities $\mathcal F_{\text{even}}$ and $\mathcal F_{\text{odd}}$ for time-frequency Gottesman–Kitaev–Preskill (GKP) states presented in the main text in Sec.~\textcolor{blue}{III.C.3}. \\ 
Consider the following wavefunction 
\begin{equation}
    \varphi_{\text{even}}(t) = \mathcal N_\text{even} e^{-\frac{t^2}{4\sigma_t^2}}\sum_{n\in \mathbb Z}g(t-2n\overline \tau),
\end{equation}
where $\mathcal N_{\text{even}}$ is a normalization constant which we choose real without loss of generality, and 
\begin{equation}
    g(t) = \frac{1}{\left(2\pi s_t^2 \right)^{1/4}}e^{-\frac{t^2}{4s_t^2}}.
\end{equation}
The Gaussian $g(t-2n\overline \tau)$ is localized around $t\approx 2n\overline \tau$. Assuming $\sigma_t \gg s_t$, the envelope value can therefore be approximated for each peak as $e^{-\frac{t^2}{4\sigma_t^2}} \approx e^{-\frac{4n^2\overline \tau^2}{4\sigma_t^2}}$.
The fidelity $\mathcal F_{\text{even}} = \left|\int dt \; \varphi_{\text{even}}(t)\varphi_{\text{even}}^*(t+\Delta x_{BA})\right|^2$ then reads 
\begin{equation}
    \mathcal F_{\text{even}} = \mathcal N_{\text{even}}^4\left|\sum_{n\in\mathbb Z}\sum_{m\in \mathbb Z}w_n w_m\int dt\; g(t-2n\overline \tau)g(t+\Delta x_{BA}-2m\overline \tau) \right|^2,
\end{equation}
where $ w_n = e^{-\frac{n^2\overline \tau^2}{\sigma_t^2}}$ and we again considered a real positive normalization constant. Rewriting the exponentials, we obtain 
\begin{equation}
    \mathcal F_{\text{even}} = \mathcal N_{\text{even}}^4 \left|\sum_{n\in \mathbb Z}\sum_{m\in \mathbb Z}w_nw_m e^{-\frac{\left(\Delta x_{BA}-2\overline \tau(m-n) \right)^2}{8s_t^2}}\right|^2,
\end{equation}
Defining $l=m-n$, 
\begin{equation}
    \mathcal F_{\text{even}} = \mathcal N_{\text{even}}^4 \left|\sum_{l \in \mathbb Z}C_l e^{-\frac{\left(\Delta x_{BA}-2\overline \tau l\right)^2}{8s_t^2}} \right|^2,
\end{equation}
where $C_l = \sum_{n\in \mathbb Z}w_n w_{n+l}$. We have 
\begin{equation}
    C_l = \sum_{n\in \mathbb Z}e^{-\frac{\overline \tau^2}{\sigma_t^2}(n^2+(n+l)^2)} =e^{-\frac{\overline \tau^2l^2}{2\sigma_t^2}}\sum_{n\in \mathbb Z}e^{-\frac{2\overline \tau^2}{\sigma_t^2}(n+l/2)^2}.
\end{equation}
In the broad-envelope limit $\sigma_t \gg \overline \tau$, we may approximate the previous sum as a Gaussian integral~\cite{TFGKPpousset2026time}, that is 
\begin{equation}
    C_l = e^{-\frac{\overline \tau^2l^2}{2\sigma_t^2}} \int dx\; e^{-\frac{2\overline \tau^2(x+l/2)^2}{\sigma_t^2}} = \frac{\sqrt{\pi}\sigma_t}{\sqrt{2}\overline \tau}e^{-\frac{\overline \tau^2l^2}{2\sigma_t^2}}.
\end{equation}
The fidelity then gives 
\begin{equation}
    \mathcal F_{\text{even}} = \mathcal N_{\text{even}}^4\frac{\pi\sigma_t^2}{2\overline \tau^2}\left|\sum_{l\in \mathbb Z}e^{-\frac{\overline \tau^2l^2}{2\sigma_t^2}} e^{-\frac{\left(\Delta x_{BA}-2\overline \tau l\right)^2}{8s_t^2}}\right|^2.
\end{equation}
The normalization constant can be computed taking $\Delta x_{BA} = 0$, as it implies a unit fidelity
\begin{equation}
    \mathcal N_{\text{even}}^{-4} = \frac{\pi\sigma_t^2}{2\overline \tau^2}\left|\sum_{l\in \mathbb Z}e^{-\frac{\overline \tau^2l^2}{2\sigma_t^2}}e^{-\frac{\overline \tau^2l^2}{2s_t^2}}\right|^2.
\end{equation}
For well separated peaks, that is for $\overline \tau \gg s_t$, the dominant contribution to the normalization is obtained for $l=0$, so that 
\begin{equation}
    \mathcal N_{\text{even}}^{-4} \approx \frac{\pi\sigma_t^2}{2\overline \tau^2}
\end{equation}
Consequently, 
\begin{equation}
     \mathcal F_{\text{even}} =  \left|\sum_{l\in \mathbb Z}e^{-\frac{\overline \tau^2l^2}{2\sigma_t^2}} e^{-\frac{\left(\Delta x_{BA}-2\overline \tau l\right)^2}{8s_t^2}}\right|^2.
\end{equation}
Likewise, the fidelity $\mathcal F_{\text{odd}}$ for odd time-frequency GKP states is evaluated by taking $2n \rightarrow 2n+1$ and $2m\rightarrow 2m+1$ in the previous calculation, and is thus equal to $\mathcal F_{\text{even}}$ to leading order in $\overline \tau/\sigma_t$ in the broad-envelope limit. The fidelity revivals, obtained for $\Delta x_{BA} = 2\overline \tau l$, have a corresponding fidelity $\mathcal F_{\text{even}} \approx e^{\frac{-\overline \tau^2l^2}{\sigma_t^2}}$.
\section{Measure-and-prepare strategy fidelity threshold derivation}
\label{Appendix Measure-and-prepare strategy fidelity threshold derivation}
In this section, we derive the measure-and-prepare strategy fidelity threshold $\max \mathcal F_{\text{class}} = 1/2$ for single-photon Gaussian wavepacket inputs presented in the main text. 
\subsection{Input Ensemble}
Consider an input ensemble $\mathcal A = \{ \ket{t_0,\omega_0;x_{\text{in}}}\}$, where $\ket{t_0,\omega_0;x_{\text{in}}}$ is a pure single-photon state,
\begin{equation}
\ket{t_0,\omega_0;x_{\text{in}}} = \int_{\mathbb R}dt\;\varphi_{t_0,\omega_0}(t-x_{\text{in}})\ket{t},
\end{equation}
with Gaussian wavepacket
\begin{equation}
\varphi_{t_0,\omega_0}(t)=\left(2\pi\sigma_t^2\right)^{-1/4}e^{-\frac{(t-t_0)^2}{4\sigma_t^2}}e^{i\omega_0(t-t_0)}.
\end{equation}
The central time and frequency parameters $(t_0,\omega_0)$ are distributed according to the Gaussian distribution
\begin{equation}
p(t_0,\omega_0)=\frac{\lambda}{2\pi}e^{-\lambda\left(\frac{t_0^2}{4\sigma_t^2}+\sigma_t^2\omega_0^2\right)},
\end{equation}
where $\lambda$ is the inverse variance and the standard deviation $\sigma_t$ is fixed throughout the ensemble $\mathcal A$. We define the central frequency with respect to an arbitrary reference frequency $\omega_{\text{ref}}$, such that $\omega_0$ is understood as a frequency detuning from $\omega_{\text{ref}}$. This allows the frequency integrals to be extended over the full real line. We consider that Alice now performs a positive operator-valued measure (POVM) constructed from the Gaussian state projectors 
\begin{equation}
\label{Equation Alice's measure-and-prepare POVM}
\hat E_{t_1,\omega_1;x_A} = \frac{1}{2\pi}\ket{t_1,\omega_1;x_A}\bra{t_1,\omega_1;x_A}.
\end{equation}
Alice measures only one photon at a given detector position, in such a way that we simply have $x_{\text{in}}=x_A$. The POVM can in principle be generalized to account for realistic detector characteristics, such as time-jitter or limited frequency bandwidth~\cite{van2017time}. \\
Alice sends the $(t_1,\omega_1)$ measurement result to Bob, who generates a new quantum state following some generation strategy $(t_1,\omega_1) \rightarrow \ket{f_{t_1,\omega_1};x_B}$. The average fidelity reads 
\begin{equation}
\begin{split}
    \mathcal F_{\text{class}}(\lambda) = \int dt_0d\omega_0\;p(t_0,\omega_0) \left(\int dt_1d\omega_1 p(t_1,\omega_1|t_0,\omega_0)|\langle f_{t_1,\omega_1};x_B| t_0,\omega_0;x_A\rangle|^2 \right),
\end{split}
\end{equation}
with $p(t_1,\omega_1|t_0,\omega_0) = \frac{1}{2\pi} \left|\langle t_1,\omega_1;x_A | t_0,\omega_0; x_A \rangle \right|^2$, the probability for the measurement result to be $(t_1,\omega_1)$ given that the input Gaussian wavepacket has $(t_0,\omega_0)$ parameters. The fidelity then yields 
\begin{equation}
    \begin{split}
        \mathcal F_{\text{class}}(\lambda) = \frac{\lambda}{4\pi^2}\int dt_0d\omega_0dt_1d\omega_1\; e^{-\frac{\left((t_0-x_A)-(t_1-x_A)\right)^2}{4\sigma_t^2}} e^{-(\omega_0-\omega_1)^2\sigma_t^2}e^{-\lambda\left(\frac{(t_0-x_A)^2}{4\sigma_t^2}+\sigma_t^2\omega_0^2\right)}
        &|\langle f_{t_1-x_A,\omega_1};x_B | t_0-x_A,\omega_0;x_A\rangle|^2,
    \end{split}
\end{equation}
shifting the time integrals by $x_A$.
\subsection{Single-Photon Gaussian Wavepackets to Coherent States}
We now introduce the correspondence $\text{TF}^{(x)}_{\alpha}: \ket{t,\omega;x}\rightarrow\ket{\alpha^{(x)}}$, which maps a pure single-photon Gaussian wavepacket parametrized by $(t,\omega)$ onto a coherent state $\ket{\alpha^{(x)}}$ with complex amplitude
\begin{equation}
\alpha^{(x)}=\frac{t-x}{2\sigma_t}+i\omega\sigma_t.
\end{equation}
The $\text{TF}^{(x)}_{\alpha}$ correspondence therefore defines a vacuum-equivalent shot-noise scale set by $\sigma_t$ and $1/(2\sigma_t)$ for the time and frequency degree of freedom, respectively. In this correspondence, the equivalent TF vacuum is represented by the Gaussian wavepacket centered at $t=x$ and $\omega=0$. As a result, the equivalent coherent-state amplitude is not an intrinsic property of the photon, but depends on the detector position $x$ that defines the reference frame. Changing the position reference from $x_A$ to $x_B$ therefore induces the transformation
\begin{equation}
\alpha^{(x_A)}\rightarrow\alpha^{(x_A)}-\delta_{BA},
\end{equation}
where $\delta_{BA}=(x_B-x_A)/(2\sigma_t)$. Since coherent states satisfy, up to a phase, $\hat D(\delta)\ket{\alpha}=\ket{\alpha+\delta}$, with $\hat D$ the displacement operator in the quadrature phase space, we obtain
\begin{equation}
\text{TF}^{(x_B)}_{\alpha}=\hat D(-\delta_{BA})\text{TF}^{(x_A)}_{\alpha}.
\end{equation}
Thus, translations of the space-time reference are represented as displacements in the effective quadrature phase space. In particular,
\begin{equation}
\text{TF}_{\alpha}^{(x_A)}\ket{t_0-x_A,\omega_0;x_A}
=\hat D\left(-\frac{x_A}{2\sigma_t}\right)\ket{\alpha},
\end{equation}
and
\begin{equation}
\text{TF}_\alpha^{(x_A)}\ket{f_{t_1-x_A,\omega_1};x_B}
=\hat D\left(-\frac{x_A}{2\sigma_t}\right)\hat D(-\delta_{BA})\ket{f_\beta},
\end{equation}
where
\begin{equation}
\alpha=\frac{t_0-x_A}{2\sigma_t}+i\omega_0\sigma_t,
\qquad
\beta=\frac{t_1-x_A}{2\sigma_t}+i\omega_1\sigma_t.
\end{equation}
 As $dt_0d\omega_0 = 2d^2\alpha$ and $dt_1d\omega_1 = 2d^2\beta$, the average fidelity is in turn equal to
\begin{equation}
\label{Equation average fidelity after Gaussian wavepacket coherent state map}
\begin{split}
    \mathcal F_{\text{class}}(\lambda) = \frac{\lambda}{\pi^2}\int d^2\beta e^{-|\beta|^2}\langle f_{\beta} |\hat O_{\beta}(\delta_{BA}) | f_\beta\rangle,
\end{split}
\end{equation}
where $\hat O_{\beta}(\delta_{BA}) = \hat D(\delta_{BA})\hat O_\beta\hat D^\dagger(\delta_{BA})$ with the positive semi-definite Hermitian operator 
\begin{equation}
\begin{split}
    \hat O_{\beta} &= \int d^2\alpha\;\ket{\alpha}\bra{\alpha}\exp(-(1+\lambda)|\alpha|^2
    +2\Re\beta^*\alpha),
\end{split}
\end{equation}
introduced in~\cite{braunstein2000criteria}, Eq.~(22). 
\subsection{Maximum Fidelity}
\label{Section Measure-and-prepare maximum fidelity}
The fidelity Eq.~\eqref{Equation average fidelity after Gaussian wavepacket coherent state map} takes the same form as in~\cite{braunstein2000criteria}, Eq.~(21), albeit with the unitary-transformed operator $\hat O_\beta(\delta_{BA})$ due to our consideration of detector-position reference frame mismatch, and accordingly 
\begin{equation}
    \mathcal F_{\text{class}}(\lambda) \leq \frac{\lambda}{\pi^2}\int d^2\beta e^{-|\beta|^2}\mu_1\left(\hat O_{\beta}\left(\delta_{BA}\right)\right),
\end{equation}
where $\mu_1\left(\hat O_{\beta}\left(\delta_{BA}\right)\right)$ is the largest eigenvalue of the $\hat O_{\beta}\left(\delta_{BA}\right)$ operator. Because $\hat O_\beta$ and $\hat O_{\beta}\left(\delta_{BA}\right)$ are related by a unitary transformation, the spectrum is preserved, and 
\begin{equation}
    \mathcal F_{\text{class}}(\lambda) \leq \frac{\lambda}{\pi^2}\int d^2\beta e^{-|\beta|^2}\mu_1\left(\hat O_{\beta}\right),
\end{equation}
as in~\cite{braunstein2000criteria}. Hence, we have $\max \mathcal F_{\text{class}}(\lambda) = \frac{1+\lambda}{2+\lambda}$, and in the limit of an ideally uniform input ensemble, that is for which $\lambda \rightarrow 0$, we have 
\begin{equation}
\label{Equation measure-and-prepare CFT}
    \max \mathcal F_{\text{class}} = \frac{1}{2}.
\end{equation}
Therefore, even though detector-position mismatch may modify the teleportation fidelity $\mathcal F_{\text{Gaussian}}$ calculated in the main text, it does not alter the optimal CFT because the corresponding benchmark operator is transformed only by unitary transformation. 
\section{Best-copy fidelity threshold derivation}
\label{Appendix Best-copy threshold derivation}
In this section, we derive the best-copy fidelity $\mathcal F_{1\rightarrow 2}$~\cite{grosshans2001quantum} for a single-photon Gaussian input, and a Gaussian cloner, discussed in Section~\textcolor{blue}{III.D.2} of the main text. \\
Consider a $1\rightarrow 2$ cloner with input mode $\text{in}$ and independent output modes $a$ and $b$. Supposing a unit gain for each channel with respective noises $\hat B$, 
\begin{align}
    \hat t_a &= \hat t_{\text{in}}+\hat B_{\hat t_a}, \; \; \hat \omega_a = \hat \omega_{\text{in}}+\hat B_{\hat \omega_a} \\
    \hat t_b &= \hat t_{\text{in}}+\hat B_{\hat t_b}, \; \; \hat \omega_b = \hat \omega_{\text{in}}+\hat B_{\hat \omega_b}.
\end{align}
Since $[\hat t_a, \hat \omega_b] = 0$, if the added noises are uncorrelated with the input, we immediately obtain $[\hat B_{\hat t_a},\hat B_{\hat \omega_b}] = -[\hat t_{\text{in}},\hat \omega_{\text{in}}]$. The Robertson inequality~\cite{hall2013quantum} then imposes 
\begin{equation}
    \Delta \hat B_{\hat t_a}\Delta \hat B_{\hat \omega_b} \geq \frac{1}{2}\left|\langle [\hat B_{\hat t_a},\hat B_{\hat \omega_b}]\rangle \right| = \frac{1}{2}\left|\langle [\hat t_{\text{in}},\hat \omega_{\text{in}}]\rangle \right| = \frac{1}{2},
\end{equation}
where we invoked the CCR $[\hat t_{\text{in}},\hat \omega_{\text{in}}] = i$ in the single-photon subspace, as shown in~\cite{TFCVfabre2022time}. Likewise, we have $\Delta \hat B_{\hat t_b}\Delta \hat B_{\hat \omega_a} \geq \frac{1}{2}$. If we assume the two copies to be identical, these two inequalities can be expressed as $\eta_t\eta_\omega \geq \frac{1}{2}$, where $\eta_t$ and $\eta_\omega$ are the time and frequency standard deviation associated with the extra noise, identical for both output modes $a$ and $b$.  \\
Let us translate this lower bound in terms of best-copy fidelity for a single-photon Gaussian input  wavepacket $ \langle t | t_0,\omega_0;x_{\text{in}}\rangle = \varphi_{t_0,\omega_0}(t-x_{\text{in}})$, where $\varphi_{t_0,\omega_0}(t)= \left(2\pi\sigma_t^2\right)^{-1/4}\;e^{-\frac{(t-t_0)^2}{4\sigma_t^2}}e^{i\omega_0(t-t_0)}$. We model the clone as a noisy copy of the input 
\begin{equation}
    \hat \rho_{\text{clone}} = \int d\delta t d\delta \omega\;p(\delta t, \delta \omega)\ket{t_0+\delta t,\omega_0+\delta \omega;x_{\text{in}}}\bra{t_0+\delta t,\omega_0+\delta \omega;x_{\text{in}}},
\end{equation}
where we assume a time-frequency noise distributed according to a Gaussian 
\begin{equation}
    p(\delta t, \delta \omega) = \frac{1}{2\pi\eta_t\eta_\omega}e^{-\left(\frac{\delta t^2}{2\eta_t^2}+\frac{\delta \omega^2}{2\eta_\omega^2} \right)}.
\end{equation}
This procedure resembles the derivation of physical time-frequency GKP states from ideal ones adopted in~\cite{fabre2020generation}, where, in time, the envelope width arises from the frequency noise and the individual peak width is due to the time noise. The fidelity between the input state and the clone is $\mathcal F_{\text{1$\rightarrow$2}} = \langle t_0,\omega_0;x_{\text{in}} | \hat \rho_{\text{clone}} | t_0,\omega_0;x_{\text{in}}\rangle$, which reads 
\begin{equation}
    \mathcal F_{\text{1$\rightarrow$2}}= \int d\delta t d\delta \omega\; p(\delta t, \delta \omega)e^{-\frac{\delta t^2}{4\sigma_t^2}}e^{-\delta \omega^2 \sigma_t^2},
\end{equation}
that is 
\begin{equation}
    \mathcal F_{\text{1$\rightarrow$2}} =  \frac{1}{\sqrt{\left(1+\frac{\eta_t^2}{2\sigma_t^2}\right)\left(1+2\sigma_t^2\eta_\omega^2\right)}}.
\end{equation}
The Robertson inequality entails $\eta_t\eta_\omega \geq 1/2$, such that the optimum is obtained for $\eta_t^2=\sigma_t^2$ and $\eta_\omega^2 = 1/(4\sigma_t^2)$ 
\begin{equation}
    \max \mathcal F_{\text{1$\rightarrow$2}} = \frac{2}{3},
\end{equation}
demonstrating the best-copy threshold discussed in the main text. Crucially, we find that the optimal time and frequency noises $\eta_t$ and $\eta_\omega$ are set by the input wavepacket standard deviation $\sigma_t$, rather than by a universal vacuum shot-noise level as in quadrature CVs, as argued in the main text. 
\bibliography{bibliocvteleportation}